\documentclass[preprint,11pt]{elsarticle}
\usepackage[letterpaper,margin=1in,includeheadfoot]{geometry}
\usepackage[T1]{fontenc}
\usepackage[utf8]{inputenc}

\usepackage{mathptmx} % consider newtxtext/newtxmath for better output

\usepackage{amsmath,amssymb,amsfonts,mathtools}
\usepackage{xfrac,bm,cancel}

\usepackage{booktabs,longtable,tabularx,array,ragged2e,multirow,hhline}
\newcolumntype{P}[1]{>{\RaggedRight\arraybackslash}p{#1}}
\newcolumntype{+}{!{\vrule width 2pt}}
\newcolumntype{b}{X}
\newcolumntype{s}{>{\hsize=.5\hsize}X}

\usepackage{float}
\usepackage{subcaption}     % <-- keep this

\usepackage{tikz}

\usepackage{xcolor}        % load BEFORE hyperref if you set link colors
\usepackage{microtype}
\usepackage{siunitx}
\usepackage[inline]{enumitem}
\usepackage{textcomp,marvosym}
\usepackage{url}           % harmless; hyperref will enhance it
\usepackage{lastpage,fancyhdr}
\usepackage{esdiff}
\usepackage{lipsum,blindtext}
\usepackage{nameref}       % okay; hyperref will patch it

\makeatletter
\newcommand{\mypm}{\mathbin{\mathpalette\@mypm\relax}}
\newcommand{\@mypm}[2]{\ooalign{%
  \raisebox{.1\height}{$#1+$}\cr
  \smash{\raisebox{-.6\height}{$#1-$}}\cr}}
\makeatother

\newlist{myenum}{enumerate}{1}
\setlist[myenum]{
    label=\arabic*.,     % 1., 2., 3.
    leftmargin=*,       % standard indentation
    align=left
}

\newenvironment{conditions*}
  {\par\vspace{\abovedisplayskip}\noindent
   \tabularx{\columnwidth}{>{$}l<{$} @{${}\quad{}$} >{\raggedright\arraybackslash}X}}
  {\endtabularx\par\vspace{\belowdisplayskip}}

\usepackage[ruled,linesnumbered,vlined,plain]{algorithm2e}
\SetArgSty{textnormal}

\usepackage[unicode,colorlinks,linkcolor=blue,citecolor=blue,urlcolor=blue]{hyperref}

\usepackage{bookmark}

\makeatletter
\newcommand{\setmaxlinesperpage}[1]{%
  \setlength{\textheight}{\dimexpr #1\baselineskip + \topskip\relax}%
}
\makeatother

\AtBeginDocument{\setmaxlinesperpage{34}}

\usepackage{setspace}
\journal{Operations Research, Data Analytics and Logistics}
\begin{document}

\begin{frontmatter}

\title{Data-Driven Bed Capacity Planning Using $M_t/G_t/\infty$ Queueing Models with an Application to Neonatal Intensive Care Units}

\author[label1,label2]{Maryam Akbari-Moghaddam\corref{cor1}}
\author[label1]{Douglas G. Down}
\author[label2,label3,label1]{Na Li}
\author[label2,label3]{Catherine Eastwood}
\author[label4]{Ayman Abou Mehrem}
\author[label4]{Alexandra Howlett}

%% Corresponding author info
\cortext[cor1]{Corresponding Author: Maryam Akbari-Moghaddam. Email: akbarimm@mcmaster.ca}

%% Affiliations
\affiliation[label1]{organization={Department of Computing and Software, McMaster University},
            addressline={1280 Main Street West},
            city={Hamilton},
            postcode={L8S 4L7}, 
            state={Ontario}, 
            country={Canada}}

\affiliation[label2]{organization={Centre for Health Informatics, Cumming School of Medicine, University of Calgary},
            addressline={3280 Hospital Drive NW},
            city={Calgary},
            postcode={T2N 4Z6},
            state={Alberta},
            country={Canada}}

\affiliation[label3]{organization={Department of Community Health Sciences, Cumming School of Medicine, University of Calgary},
            addressline={3280 Hospital Drive NW},
            city={Calgary},
            postcode={T2N 4Z6},
            state={Alberta},
            country={Canada}}

\affiliation[label4]{organization={Department of Pediatrics, Cumming School of Medicine, University of Calgary},
            addressline={Alberta Children’s Hospital, 28 Oki Drive NW},
            city={Calgary},
            postcode={T3B 6A8},
            state={Alberta},
            country={Canada}}

%% Abstract

\nonumnote{This paper has been submitted to the 
\textit{Operations Research, Data Analytics and Logistics} journal.}

\begin{abstract}

Hospitals face challenges in long-term intensive care unit (ICU) capacity planning under uncertain demand. Admission rates fluctuate over time, and LOS distributions vary with patient heterogeneity, hospital location, case mix, and clinical practice. Common approaches rely on steady-state queueing models or heuristic rules with fixed parameters, which often fail to capture real occupancy dynamics. The widely used 85\% occupancy rule, for example, recommends keeping average utilization below this level to preserve responsiveness, yet it is grounded in stationary assumptions and may lack resilience in time-varying systems. Our analysis shows that even when long-run utilization targets are satisfied, daily occupancy often exceeds 100\% capacity.

We propose a data-driven framework to estimate ICU bed occupancy using an $M_t/G_t/\infty$ queueing model with time-varying arrival rates and empirically fitted LOS distributions. The approach combines statistical decomposition and parametric fitting to capture temporal patterns in admissions and LOS, and is applied to multi-year data from neonatal ICUs (NICUs) in Calgary. We evaluate capacity scenarios including average-based thresholds and Poisson-based surge estimates. Results show that static heuristics are inadequate under fluctuating demand and underscore the importance of modeling LOS variability when estimating bed needs. Although the case study focuses on NICUs, the framework has potential applicability to other ICU settings and provides interpretable, data-informed support for systems facing rising demand and constrained capacity.
\end{abstract}

\begin{keyword}
Bed occupancy forecasting 
\sep Data-driven planning
\sep Time-varying queueing models
\sep Length of stay (LOS) modeling
\sep Capacity planning
\sep $M_t/G_t/\infty$ queue
\sep Intensive Care Units (ICUs)

\end{keyword}

\end{frontmatter}

\section{Introduction} \label{Intro}

Intensive care units (ICUs) provide continuous monitoring and advanced interventions for critically ill patients. Their capacity is limited by staffed beds, specialized personnel, and infrastructure. When demand approaches or exceeds available resources, care delivery can become strained, which reflects a supply–demand mismatch that affects beds, staffing, and equipment \citep{wunsch2009use, rewa2018indicators, batchelor2021adult}. Many health systems are planning for sustained pressure in critical care while facing constraints such as staffing shortages and aging infrastructure \citep{european2021state, murthy2015intensive}. These challenges make long-term capacity planning a central operational problem in ICU management.

A key difficulty is that ICU demand and service durations vary over time. Admissions fluctuate due to temporal factors, seasonal illness patterns, and population dynamics \citep{whitt2019forecasting, eick1993mt}. Length of stay (LOS) varies widely because of heterogeneity in patient severity and comorbidities \citep{kim1999analysis}, and can evolve with changes in clinical practice, case mix, and hospital protocols. Because occupancy depends on both the arrival process and the full LOS distribution, planning based on time-averaged quantities can mask periods of congestion and yield capacity targets that lack resilience during peak periods.

Traditionally, long-term ICU bed planning has relied on steady-state queueing models that assume stationary arrival rates and service durations \citep{helm2011design}. These models use long-run average admission rates and LOS, which simplifies analysis and remains common in practice due to tractability \citep{green2007coping}. However, when arrivals and discharges vary over time, as often observed in ICUs because of seasonal patterns or structural demand shifts, stationary assumptions fail to capture real dynamics \citep{eick1993mt, chan2017queues}. As a result, steady-state models may underestimate capacity during high-demand periods \citep{green2001strategies} or encourage overprovisioning when planning relies on historical peaks. Moreover, occupancy depends on the full LOS distribution rather than its mean, since LOS variability determines how long patients overlap in the system.

This study develops a data-driven framework for long-term ICU bed planning under non-stationary demand. We focus on neonatal intensive care units (NICUs) across five hospital sites in the Calgary Zone, Alberta, Canada, which exhibit temporal variation in admissions and LOS. These sites operate within a regional network where patient transfers often respond to occupancy levels. Our goal is to estimate the beds required if each site accommodated its underlying demand without capacity-driven redistribution.

Although NICUs have distinct clinical features, they share key capacity constraints with other ICUs, including limited flexibility in staffed beds and the need to maintain readiness under demand uncertainty \citep{kim1999analysis}. Infants requiring advanced support often need immediate admission, even when the unit is strained. Congestion therefore appears as overload and resource strain rather than infants waiting for resources. This characteristic motivates the use of an $M_t/G_t/\infty$ framework, an infinite-server queueing model that accommodates time-varying arrivals and service durations rather than relying on steady-state averages. The infinite-server assumption reflects the priority of immediate admission in critical care. Strain is therefore expressed as overload and resource bottlenecks rather than waiting lines. The model yields a tractable expression for expected occupancy via convolution of past arrivals with time-varying LOS survival probabilities, enabling transparent scenario evaluation.

Our contributions are as follows. First, we estimate time-varying admission rates and mean LOS using STL decomposition \citep{rb1990stl}, selected through a grid search. Second, we fit parametric LOS distributions and construct time-varying survival probabilities by calibrating parameters to smoothed daily LOS moments while preserving site-level tail heterogeneity. Third, we integrate these components in a non-stationary model to generate daily occupancy trajectories under the $M_t/G_t/\infty$ framework. Fourth, we convert occupancy into capacity targets using benchmarks ranging from average-based rules to overflow probability constraints based on Poisson tail approximations. Finally, we extend the model to forward-looking planning through a births-driven projection module combining demographic forecasts with resampled historical arrival and LOS patterns.

Our framework produces site-specific capacity estimates that reflect temporal variability in demand and LOS. The results show that targets based solely on time averages may fail to protect against transient overload, while overflow-constrained thresholds provide explicit control of exceedance risk at the cost of lower utilization. The projection module generates scenario-based ranges for future bed requirements under demographic change.

The remainder of the paper is organized as follows. Section \ref{LitReview} reviews related work on capacity heuristics, queueing models, and forecasting approaches for hospital occupancy. Section \ref{ProblemData} describes the Calgary NICU dataset and the planning objective. Section \ref{MatMet} presents the modeling framework, including STL-based estimation, LOS distribution fitting, and occupancy measures derived from the queueing model. It also defines the capacity planning strategies and resilience metrics. Finally, Section \ref{Results} reports empirical results and scenario comparisons. Section \ref{Discussion} summarizes implications, limitations, and extensions for ICU capacity planning.

\section{Related Work} \label{LitReview}

The estimation of hospital bed occupancy has been studied using various modeling paradigms. This section reviews prior work in three areas relevant to our study: (i) queueing models and capacity planning frameworks, (ii) simulation-based forecasting, and (iii) time-series and machine learning forecasting approaches.We highlight how this literature informs our framework and how our model extends existing work to support data-driven, scenario-based ICU capacity planning. Although similar forecasting and capacity planning frameworks have been applied in other operational domains, we focus on healthcare-specific literature due to the distinct clinical and patient-flow constraints inherent to healthcare systems.

\subsection{Queueing Models and Capacity Planning Frameworks}

A widely used approach in hospital planning is steady-state queueing models, particularly $M/M/c$ or Erlang-based frameworks, which assume constant arrival and service rates to estimate required bed capacity. These models are widely applied in healthcare resource planning \citep{gorunescu2002queueing, gorunescu2002using, pinto2014analisys,shapoval2017optimizing, bittencourt2018hospital}. Some studies embed them within cost-optimization frameworks to determine hospital-specific occupancy and capacity targets \citep{maaz2020determining}. Others, such as Joseph \cite{joseph2020queuing}, stress their practical value while cautioning against simplistic average-based decisions.

Closely related to steady-state reasoning is the 85\% rule, which suggests that average occupancy in high-resilience settings such as ICUs and emergency departments should not exceed 85\% \citep{janke2022hospital}. The rule aims to ensure sufficient capacity for demand surges and reduce delays but is grounded in long-run averages. It does not explicitly account for temporal variability in arrivals or service patterns. Bain et al.\ \cite{bain2010myths} argue that a universal ``safe'' occupancy threshold oversimplifies queueing dynamics and obscures the trade-off between utilization and availability, emphasizing the need for system-specific evaluation rather than fixed benchmarks.

These limitations are amplified in NICUs, where admissions and LOS vary substantially and flexibility in staffed beds is limited. In such settings, steady-state models and fixed occupancy rules fail to capture dynamic demand and short-term overload risk. Several studies propose time-varying congestion metrics beyond long-run averages. Au et al.\ \cite{au2009predicting} model emergency department overflow as a time-dependent queue and show that congestion depends on patient flow and bed turnover dynamics. Wartelle et al.\ \cite{wartelle2024changing} introduce a time-windowed congestion metric based on arrival-to-departure load ratios, arguing that averages such as occupancy and waiting time do not reflect short-term congestion patterns.

Despite the relevance of non-stationary queueing to hospital operations, many applied planning studies still rely on steady-state approximations using time-averaged arrivals and LOS \citep{helm2011design, green2007coping}. Some work adopts dynamic queueing models to better reflect variability. Eick et al.\ \cite{eick1993mt} analyze the $M_t/G/\infty$ queue under sinusoidal input, and Whitt and Zhang \cite{whitt2019forecasting} integrate SARIMA-based arrivals with deterministic queueing for ED forecasting. Other studies extend the $M_t/G/\infty$ framework to address data or system constraints. For example, Li et al.\ \cite{li2019statistical} estimate parameters from interval-censored data via maximum likelihood, Whitt and Zhao \cite{whitt2017many} incorporate non-Poisson arrivals with Gaussian staffing approximations, and Chan et al.\ \cite{chan2017queues} model inspection-based discharge timing. Shi et al.\ \cite{shi2016models} propose a broader processing network capturing inpatient delays and overflow policies.

\subsection{Simulation-based Forecasting} \label{twotwo}

Several papers propose data-driven models for short-term occupancy forecasting. Baas et al.\ \cite{baas2021real} use Monte Carlo simulation of patient trajectories through wards and ICUs during COVID-19 to generate real-time bed forecasts, combining empirical LOS with epidemic-based arrival projections. Similarly, Whitt and Zhang \cite{whitt2017data} apply a time-varying infinite-server model to emergency department data to capture weekly patterns in arrivals, admissions, and LOS. Their focus is short-term operations, whereas our objective is long-term capacity planning. Leeftink et al.\ \cite{leeftink2025inter} simulate nurse shift allocation across a national NICU network to address daily demand fluctuations. Braaksma et al.\ \cite{braaksma2021bed} use hourly census forecasts to guide intraday nurse staffing and workload balancing through flexible float deployment.

Other simulation frameworks address multi-layered or regionally coordinated planning. Dijkstra et al.\ \cite{dijkstra2023dynamic} develop a multi-level simulation-optimization model to redistribute COVID-19 patients across hospitals, combining infinite-server queues with stochastic programming to balance regional surpluses and shortages. However, these approaches require centralized coordination and real-time transfer control, which are not available in our NICU setting and fall outside this study's scope.

\subsection{Time-Series and Machine Learning Forecasting Approaches} \label{twothree}

Parallel to queueing approaches, a large literature addresses occupancy and demand forecasting using statistical and machine learning methods. Classical time-series models such as SARIMA \citep{box2015time} and more recent tools such as Prophet \citep{taylor2018forecasting} and XGBoost \citep{chen2016xgboost} have been widely used in healthcare forecasting.

Tuominen et al.\ \cite{tuominen2021forecasting} forecast daily arrivals and use peak occupancy as a proxy for crowding in a Finnish ED. Cheng et al.\ \cite{cheng2021forecasting} and Reboredo et al.\ \cite{reboredo2023forecasting} apply SARIMAX and INGARCH models to short-term ED forecasting. The former predicts hourly occupancy up to four hours ahead, while the latter forecasts daily arrivals for staffing decisions. Both account for temporal autocorrelation, and INGARCH also captures time-varying variability in arrivals.

Recent studies examine richer feature sets and explainable AI methods. Tuominen et al.\ \cite{tuominen2024forecasting} evaluate LightGBM, N-BEATS, DeepAR, and TFT for 24-hour ED occupancy forecasting using over 150 covariates, finding that simpler models can match or outperform deep learning benchmarks such as ARIMA. Susnjak and Maddigan \cite{susnjak2023forecasting} apply ensemble learning with SHAP and LIME for three-month urgent care forecasts. Becerra et al.\ \cite{becerra2020forecasting} use SARIMA to model seasonal respiratory ED visits in Chile. Overton et al.\ \cite{overton2022epibeds} propose EpiBeds, linking SEIR projections with hospital transition pathways to produce weekly ICU demand forecasts in the United Kingdom.

While these forecasting approaches support operational prediction, they often omit explicit LOS distributions and lack tractable links between demand, LOS, and long-term capacity decisions. In domains such as call centers, staffing can be adjusted flexibly in response to demand \citep{gans2003telephone, whitt2006staffing, koole2023practice}. ICU bed capacity, however, is constrained by infrastructure and specialized staffing, motivating planning tools that convert time-varying demand and service patterns into interpretable capacity thresholds.

Finally, Chen et al.\ \cite{chen2024can} study sinusoidal non-homogeneous Poisson processes (NHPPs) for modeling arrivals in customer service systems, showing they capture non-weekly cycles more smoothly than piecewise linear functions. Although we estimate arrival rates through empirical decomposition, this work supports modeling time-varying demand with flexible functional forms.

Within this literature, few models integrate time-varying arrival rates and LOS distributions in a data-driven queueing framework for long-term planning. Most emphasize short-term forecasting without scenario analysis or rely on simulation without tractable structure. Many studies in Sections \ref{twotwo} and \ref{twothree} focus on demand prediction, staffing optimization, or patient flow rather than physical bed capacity. Empirical estimation of LOS variance and its effect on occupancy uncertainty is also often overlooked despite its relevance for surge planning. Our approach addresses these gaps by combining statistical decomposition, parametric LOS modeling, and occupancy estimation within an infinite-server framework to generate interpretable, site-specific planning thresholds.

\section{Problem Setting and Data} \label{ProblemData}

We begin by describing the general modeling framework and then present the empirical setting used for implementation and evaluation. This study develops a data-driven approach to estimate time-varying bed occupancy and the capacity required to meet specified resiliency targets when arrivals and service durations evolve over time. Planning in this setting is challenging because admissions and LOS fluctuate, while capacity cannot be scaled quickly in response to surges. As a result, average-based rules can be misleading, and explicit quantification of occupancy variability is essential for robust long-term planning.

\subsection{Modeling Objective}

The objective of this study is twofold. Methodologically, we estimate daily expected bed occupancy in a critical care unit using empirically observed, time-varying arrival rates and LOS distributions. Expected occupancy varies with seasonal admission patterns and evolving clinical practices that affect average stay duration. Because daily occupancy is stochastic and fluctuates beyond average trends, planning cannot rely solely on long-run averages.

Second, we translate these stochastic occupancy estimates into capacity requirements that satisfy explicit probabilistic resilience targets. We estimate the probability that occupied beds exceed available capacity and use it to determine capacity levels under specified resiliency thresholds. Capacity is set as the minimum number of beds needed to keep this exceedance probability below a chosen limit. This framework also enables comparison of alternative capacity levels based on their implied overflow risk over time.

\subsection{Data Description}

We utilize a multi-year dataset of NICU admissions from five major sites in the Calgary Zone between April 1, 2016 and December 31, 2023, with each record representing one admission. The data include patient demographics, admission characteristics, diagnosis codes, and timestamps, enabling reconstruction of daily census counts and estimation of site-specific demand and LOS patterns. For modeling, we use only the institution identifier and total NICU LOS in days, treated as continuous to capture partial-day stays. These variables are sufficient to estimate time-varying occupancy within the queueing framework.

\begin{figure}[ht]
    \centering
    \begin{subfigure}[t]{0.49\textwidth}
        \centering
        \includegraphics[width=\textwidth]{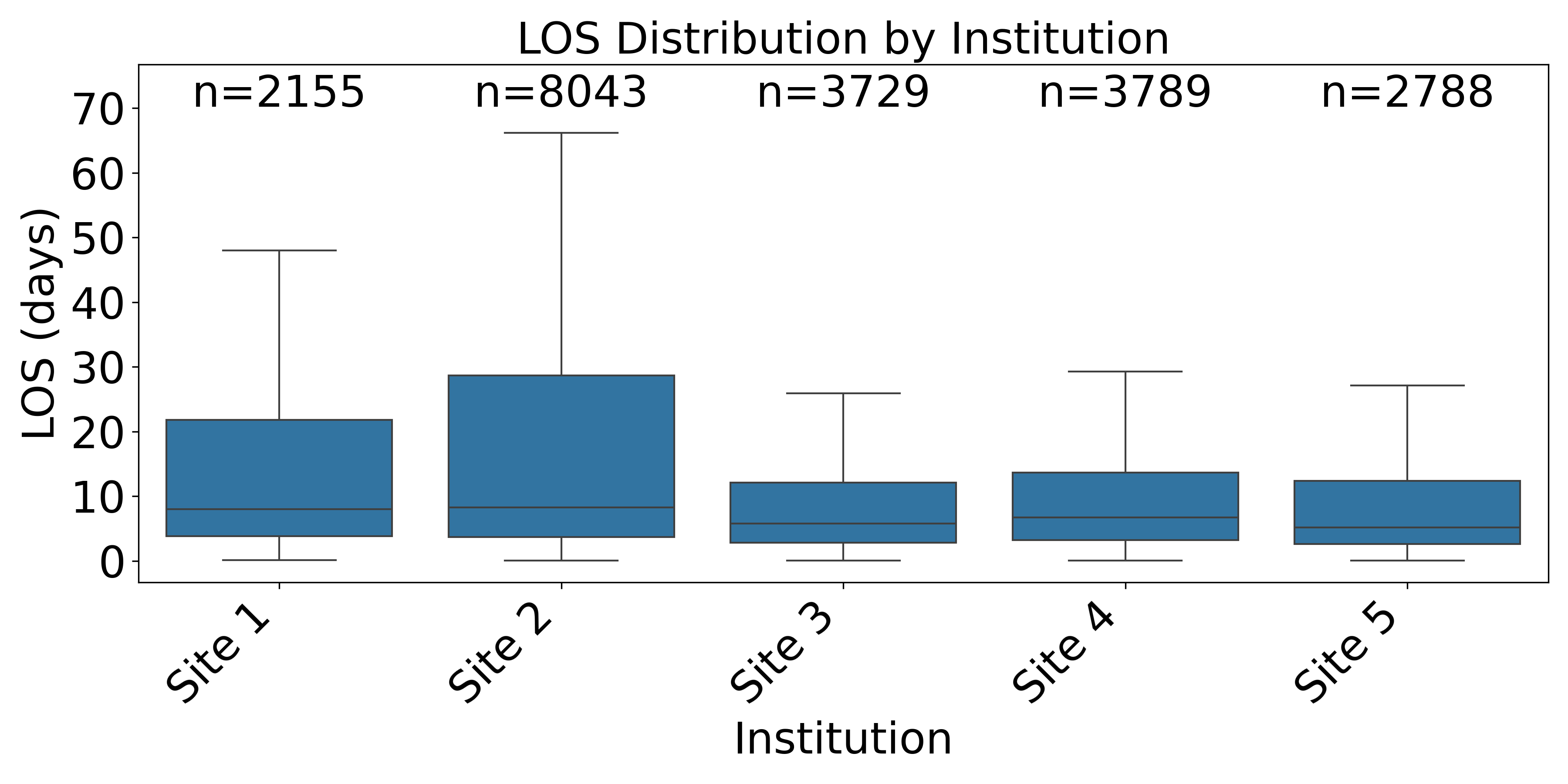}
        \caption{LOS distribution by institution (adjusted)}
        \label{fig:los_by_inst}
    \end{subfigure}
    \hfill
    \begin{subfigure}[t]{0.49\textwidth}
        \centering
        \includegraphics[width=\textwidth]{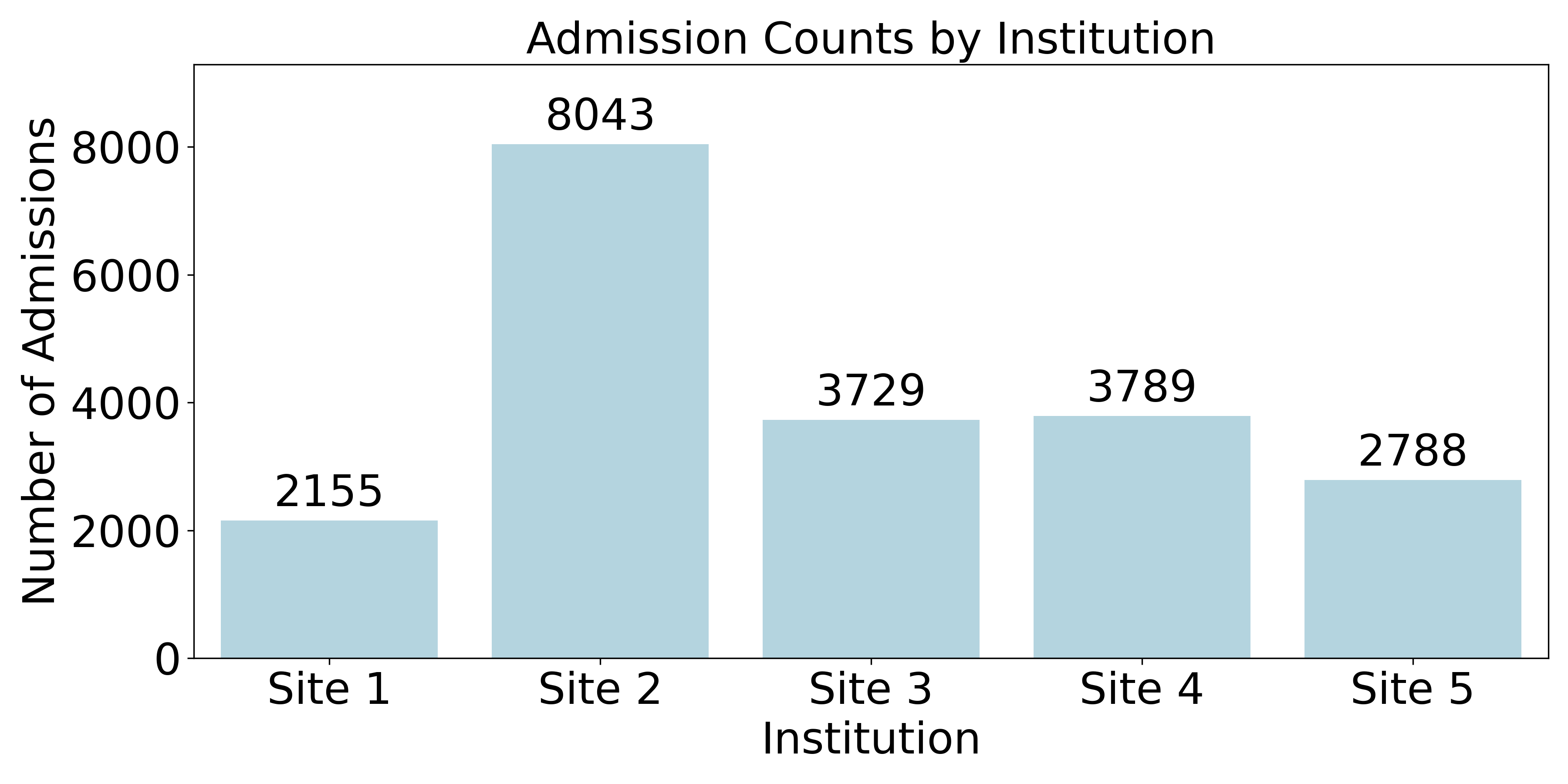}
        \caption{Admission counts by institution (adjusted)}
        \label{fig:admits_by_inst}
    \end{subfigure}
    \caption{Comparison of adjusted LOS distribution and admission counts by institution using adjusted series that retain capacity-driven transfers at the originating site.}
    \label{fig:los_admits_by_inst}
\end{figure}

A key feature of the Calgary Zone NICU system is that patient transfers between sites are often capacity driven. We focus on estimating the number of beds required if each site accommodated its own demand. Therefore, we construct adjusted site-level demand and LOS time series before modeling. Specifically, for Site~2, many admissions historically transfer to Sites 3–5 during periods of high congestion. In the adjusted dataset, these admissions and their LOS are attributed back to Site~2 rather than the receiving sites. Consequently, adjusted admissions at Sites 3–5 decline, while Site~2’s LOS increases. Figure~\ref{fig:los_admits_by_inst} shows the adjusted LOS distributions and volumes. All subsequent analyses use these adjusted series.

\subsection{Challenges in Modeling NICU Bed Occupancy}

Estimating NICU occupancy requires modeling both the arrival process and LOS distribution, both of which exhibit heterogeneity and non-stationarity. Admission volumes and LOS differ substantially across sites, reflecting variation in unit size, clinical practice, patient mix, and historical transfer patterns.

\begin{figure}[ht]
\centering
\begin{subfigure}[t]{0.49\textwidth}
    \centering
    \includegraphics[width=\textwidth]{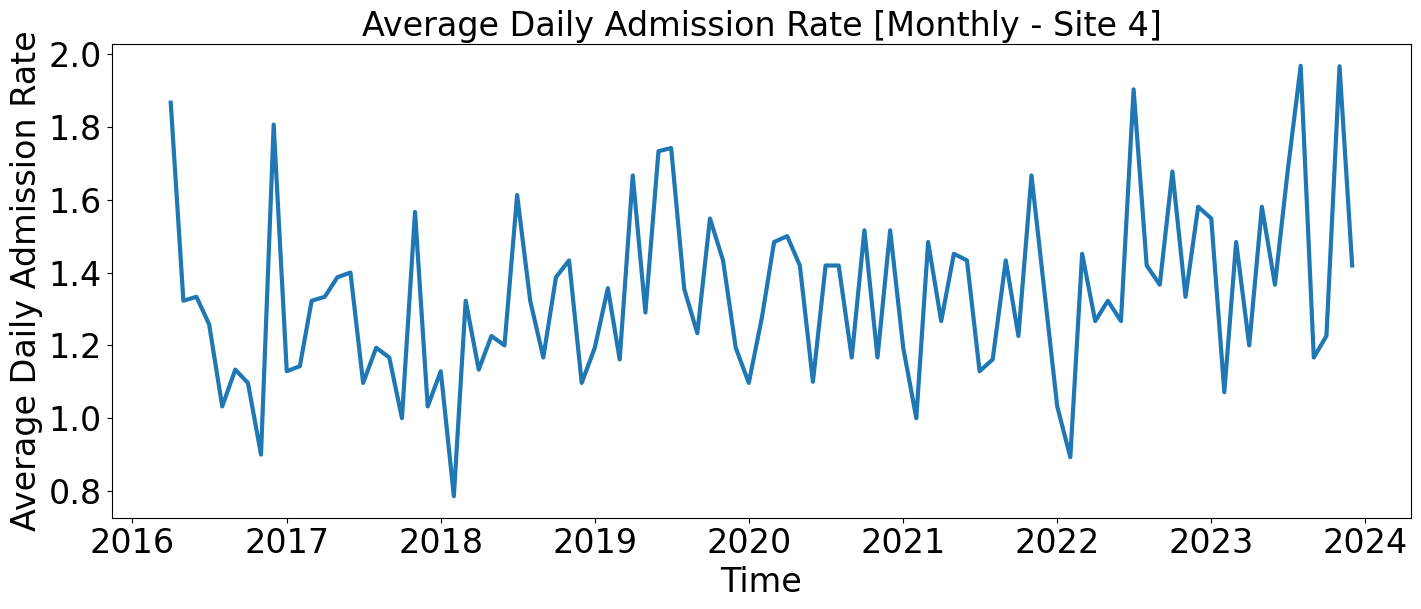}
    \caption{Average daily admission rate by month (adjusted)}
    \label{fig:lambda_over_time_80148}
\end{subfigure}
\hfill
\begin{subfigure}[t]{0.49\textwidth}
    \centering
    \includegraphics[width=\textwidth]{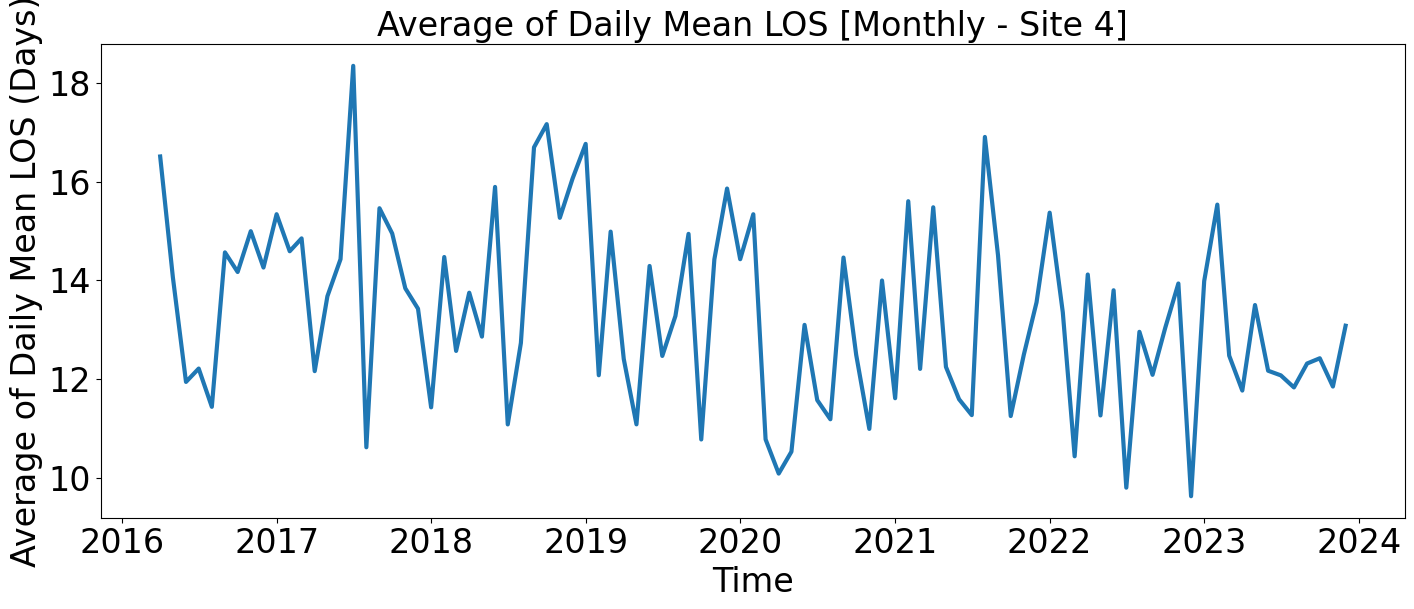}
    \caption{Average of daily mean LOS by month (adjusted)}
    \label{fig:mean_los_over_time_80148}
\end{subfigure}
\caption{Time-varying admissions and mean LOS at an example site using the adjusted series.}
\label{fig:lambda_los_trends}
\end{figure}

Figure \ref{fig:los_admits_by_inst} shows substantial heterogeneity in LOS distributions and admission volumes across the five sites, which motivates site-specific estimation of arrival rates and LOS. In addition to cross-site heterogeneity, both admissions and LOS vary over time. Admissions display seasonality and short-term fluctuations linked to birth patterns, population dynamics, and operational changes. LOS is also non-stationary, with its mean and variability shifting due to evolving clinical protocols or case mix. Figure~\ref{fig:lambda_los_trends} illustrates these dynamics for one representative site, with similar patterns observed across others.

To determine whether these visual patterns reflect significant changes over time, we conducted site-level trend analyses for monthly admissions and mean LOS. For each NICU, the monthly series was regressed on a linear time index with month-of-year indicators, using Newey–West (HAC) robust standard errors to account for serial correlation. Results show statistically significant trends in admissions and mean LOS at most sites, indicating that the patterns in Figure~\ref{fig:lambda_los_trends} represent systematic changes rather than random month-to-month variation.

Finally, the multi-site structure introduces an additional consideration for capacity planning. Because patient transfers redistribute load, site-level capacity estimates depend on whether the system is treated as independent units or a connected network. We estimate requirements using adjusted demand that attributes capacity-driven transfers back to the originating site. This choice aligns the modeled demand with the capacity that would be needed if reliance on redistribution were reduced, and it gives insights into how transfer policies can affect the apparent balance of demand and capacity across sites.

\section{Methodology} \label{MatMet}

This section describes the framework used to estimate time-varying occupancy and derive capacity under resiliency constraints. It integrates empirical estimation of arrival rates and LOS distributions within a non-stationary infinite-server queueing model to support scenario-based planning across ICU sites facing demand uncertainty and service variability.

\subsection{Overall Framework} \label{overallframework}

Our framework estimates and evaluates time-varying ICU occupancy under non-stationary conditions using an $M_t/G_t/\infty$ model. It integrates empirical estimation of the arrival process and LOS distribution and supports scenario-based capacity planning. The main components are: (i) arrival rate estimation, (ii) LOS distribution modeling, and (iii) estimation of expected occupancy under the infinite-server queueing model. Each component is implemented in a site-specific manner to reflect operational heterogeneity across ICUs. Table~\ref{tab:notation} summarizes the key indices, variables, and planning parameters.

\begin{table}[!ht]
\centering
\footnotesize
\setlength{\tabcolsep}{4pt}
\renewcommand{\arraystretch}{0.72}
\caption{Notation used throughout the paper}
\label{tab:notation}
\begin{tabular}{p{2cm}p{8cm}}
\toprule
Symbol & Definition \\
\midrule

$m$ & Index for NICU site \\
$t$ & Day index \\
$u$ & Days since admission, used as the convolution lag \\
$S$ & LOS random variable \\
$S_{\max}$ & Truncation horizon in the convolution \\
$\lambda_t$ & Time-varying admission rate on day $t$ \\
$\mu_t$ & Time-varying mean LOS on day $t$ \\
$\sigma_t^2$ & Time-varying LOS variance on day $t$ \\
$\kappa$ & LOS shape parameter \\
$\rho_t$ & Expected occupancy on day $t$ \\
$L_t$ & Number of occupied beds on day $t$ \\
$\bar{\rho}$ & Long-run average occupancy implied by historical averages \\
$\delta_t$ & Deviation from long-run mean occupancy\\
$B_{\text{average}}$ & Baseline capacity estimate using historical averages \\
$B_{\alpha}$ & Capacity level corresponding to overflow risk threshold $\alpha$ \\
$\gamma$ & Utilization threshold used in overflow evaluation \\
$\eta$ & Elasticity of admissions with respect to projected births \\
$\psi$ & Multiplicative drift factor in projections \\
$y$ & Projection year index \\
$Y_f$ & Set of projection years, $Y_f=\{y_{\min},\ldots,y_{\max}\}$ \\
$Y_h^{\omega}$ & Recent-year reference set used to define baseline admissions and site shares \\
$Y_h^{\nu}$ & Historical reference set used to resample within-year arrival and LOS patterns \\
$h, h'$ & Resampled reference years drawn from $Y_h^{\nu}$ for arrivals and LOS, respectively \\
$A_y$ & Observed total admissions in year $y$ \\
$\hat{A}_y$ & Projected system-wide admissions in year $y$ \\
$\tilde{K}_y$ & Projected number of births in year $y$ \\
$s_m$ & Average admission share of site $m$ over $Y_h^{\omega}$ \\

\bottomrule
\end{tabular}
\end{table}

Because NICU discharge decisions may be influenced by congestion, we tested whether LOS was associated with contemporaneous occupancy levels. Cox models at pooled and site-specific levels showed no significant link between occupancy levels and time to discharge. We therefore treat LOS as independent of occupancy levels. In settings where such effects exist, LOS would need to be modeled as state-dependent, requiring extension of the $M_t/G_t/\infty$ framework to allow LOS to depend on the occupancy level.

We formulate the problem as follows. Let $\lambda_t$ denote the admission rate on day $t$, allowed to vary over time. The expected occupancy on day $t$, denoted $\rho_t$, accounts for patients admitted on prior days $t-u$ whose LOS extends beyond $t$, where $u \in \{0,1,\ldots,S_{\max}\}$ and $S_{\max}$ is a truncation horizon. This requires the conditional survival probability $
\mathbb{P}(S_k > u \mid \tau_k = t-u) = 1 - G_{t-u}(u)$,
the probability that patient $k$, admitted on day $\tau_k = t-u$, remains hospitalized on day $t$. Here, $S_k$ denotes the LOS of patient $k$, and $G_{t-u}$ is the cumulative distribution function (CDF) of the LOS distribution for admissions on day $t-u$, estimated from best-fitting parametric models with time-varying parameters. Although indexed by $k$, we assume a common LOS distribution across patients, so the survival probability depends only on time since admission and the subscript $k$ is suppressed in implementation.

This survival probability does not treat death as a competing risk. LOS represents total time to exit, whether by discharge, transfer, or death, since occupancy modeling concerns only when a bed becomes available. The queueing model therefore treats LOS as time to system exit without distinguishing exit mechanisms. We next describe the key components in detail.

\subsection{Arrival Rate Estimation}

To estimate the time-varying arrival rate $\lambda_t$ for each site $m$, we use historical admission data aggregated into daily counts from the adjusted series described in Section~\ref{ProblemData}. Because admissions exhibit noise, seasonality, and irregular fluctuations, we apply statistical decomposition to extract a smooth estimate of the underlying arrival intensity for the occupancy model.

Specifically, we apply Seasonal-Trend decomposition via Loess (STL) \citep{rb1990stl} to decompose daily admissions into trend, seasonal components, and residual noise. STL is chosen for robustness to outliers and flexibility in capturing irregular seasonality. For each site, we perform a grid search over seasonal windows of 7, 15, and 31 days, trend windows of 15, 31, and 61 days, polynomial degrees of 1 and 2, and a robustness flag for weighted fitting. These windows reflect weekly to monthly smoothing scales relevant to hospital admissions. The grid search selects the configuration that minimizes residual standard deviation. Our search evaluates 72 candidate configurations per site. The resulting smoothed trend is used as the estimate of $\lambda_t$, representing expected admissions at site $m$ on day $t$.

We use STL over alternatives like SARIMA \citep{box2015time} or machine learning because our goal is not short-term forecasting but estimation of a smooth, interpretable arrival function for queueing analysis. Unlike black-box models, STL offers explicit separation of trend and seasonality and supports scenario evaluation \citep{hyndman2018forecasting}. Moreover, its outputs can be directly incorporated into the $M_t/G_t/\infty$ framework.

\subsection{LOS Distribution Modeling} \label{LOSmodeling}

To estimate the time-varying conditional survival probability $\mathbb{P}(S > u \mid \tau = t - u)$ for each site $m$ and day $t$, we use historical LOS data for all admissions. The same preprocessing and STL decomposition applied to admission counts is used for LOS, retaining the grid search configuration. LOS values are aggregated into daily averages, and STL extracts a smooth trend that estimates the time-varying mean LOS \( \mu_t \). To capture dispersion, we analyze STL residuals with rolling windows of 7, 15, and 31 days and select the window yielding the most stable local volatility. The rolling standard deviation is then squared to obtain the time-varying LOS variance \( \sigma_t^2 \).

We then fit several candidate parametric distributions to empirical LOS values at each site, including Weibull, Lognormal, Gamma, Fisk (Burr Type XII), and Exponential families. These distributions are commonly used for neonatal LOS due to their ability to capture skewness and heterogeneity \citep{kheiry2020evaluation}. For each site, maximum likelihood estimation (MLE) is applied to pooled LOS observations to obtain time-invariant distributional parameters, independently of arrivals and time-varying LOS moments.

Based on empirical analysis, we observed that the shape parameter $\kappa$ of the Weibull, Gamma, and Fisk distributions remains stable over time when LOS is aggregated at quarterly, biannual, and annual resolutions. We assess this using the coefficient of variation (CV) of $\kappa$ across aggregation windows and find CV < 0.2 in all cases \citep{shechtman2013coefficient}. We therefore fix $\kappa$ per site and deterministically re-parameterize the remaining distribution parameters to match the smoothed daily LOS mean and, when applicable, variance obtained from STL decomposition. These time-varying moments are not used in the MLE stage but are incorporated when constructing conditional survival probabilities for the occupancy model. While this approach is supported by the stability observed in our data, in general, one may need to estimate a time-varying function for the shape parameter to account for potential structural changes or temporal dynamics.

To evaluate goodness-of-fit, we compare the marginal survival function of each candidate distribution to the empirical survival curve from the Kaplan-Meier estimator. We compute the root mean squared error (RMSE) between empirical and parametric survival probabilities $\mathbb{P}(S > u)$ over a distribution-specific horizon $S_{\max}$. For each distribution, $S_{\max}$ is defined as the smaller of the maximum observed LOS at that site and the 99th percentile of the corresponding parametric distribution constructed using fixed shape parameters (when applicable), the empirical mean LOS, and variance (when required). This ensures comparison over a range where both empirical and model-based survival probabilities are well defined. The distribution minimizing RMSE is selected as best-fitting for that site. The selected distribution, with fixed $\kappa$ (if applicable), is then used to compute conditional survival probabilities $\mathbb{P}(S > u \mid \tau = t - u)$ in Eq. \ref{eq:1} below.

The conditional survival probabilities $\mathbb{P}(S > u \mid \tau = t - u)$ are defined as follows for each candidate distribution:

\begin{itemize}

    \item \textbf{Exponential:} \quad \(\exp\left( -\frac{u}{\mu_{t - u}} \right)\),

  \item \textbf{Weibull:} \quad \(\exp\left( -\left( \frac{u}{\theta_{t - u}} \right)^\kappa \right)\), \quad where \(\theta_{t - u} = \frac{\mu_{t - u}}{\Gamma\left(1 + \frac{1}{\kappa}\right)}\),

  \item \textbf{Lognormal:} \quad \(1 - \Phi\left( \frac{\log u - \mu^{\text{lognorm}}_{t - u}}{\tau_{t - u}} \right)\), \quad where \(\tau_{t - u} = \sqrt{ \log\left(1 + \frac{\sigma^2_{t - u}}{\mu^2_{t - u}} \right)}\), \quad and \(\mu^{\text{lognorm}}_{t - u} = \log(\mu_{t - u}) - \frac{1}{2} \tau^2_{t - u}\),

  \item \textbf{Gamma:} \quad \(1 - F_{\text{Gamma}}(u; \kappa, \theta_{t - u})\), \quad where \(\theta_{t - u} = \frac{\mu_{t - u}}{\kappa}\), and \(F_{\text{Gamma}}(u; \kappa, \theta)\) denotes the CDF of the Gamma distribution with shape parameter \(\kappa\) and scale parameter \(\theta\), evaluated at \(u\),

    \item \textbf{Fisk (Burr Type XII):} \quad \(\left( \frac{\theta_{t - u}}{u + \theta_{t - u}} \right)^\kappa\), \quad where \(\theta_{t - u} = \frac{\mu_{t - u}}{\pi / \kappa}\) (if \(\kappa \neq 0\)).

\end{itemize}

In Section \ref{OccupancyEstimation}, we discuss how we estimate site-specific bed occupancy and provide several scenario-based capacity planning strategies.

\subsection{Occupancy Estimation and Scenario-Based Capacity Planning} \label{OccupancyEstimation}

Given the time-varying admission rate $\lambda_t$ and the conditional survival probability function $\mathbb{P}(S > u \mid \tau = t - u)$, we estimate the expected beds occupied at each ICU site $m$ on day $t$ using a non-stationary infinite-server queue. As discussed in Section~\ref{overallframework}, this is modeled under the $M_t/G_t/\infty$ framework, which allows both the arrival rate and LOS distribution to vary over time.

The expected occupancy $\rho_t$ is computed as a convolution of past admissions with corresponding survival probabilities. For each day $t$, we sum patients admitted on days $t - u$, for $u = 0$ to $S_{\max}$, weighting each by the probability of remaining hospitalized on day $t$. The expected bed occupancy is calculated using the standard convolution formula for infinite-server queues:

\begin{equation}
\rho_t = \sum_{u=0}^{S_{\max}} \lambda_{t - u} \cdot \mathbb{P}(S > u \mid \tau = t - u).
\label{eq:1}
\end{equation}

We may interpret the expected occupancy $\rho_t$ through the decomposition $
\rho_t = \bar{\rho} + \delta_t$,
where $\bar{\rho}$ represents occupancy implied by long-run average arrival and LOS rates, and $\delta_t$ captures excess occupancy from short-term fluctuations (due to both LOS variance and temporal fluctuations). Under this view, capacity planning consists of providing beds for long-term demand plus a buffer to absorb day-to-day short-term fluctuations.

Eq. \ref{eq:1} builds on the classical \( M_t/G/\infty \) result of Eick et al.\ \cite{eick1993mt} and its extension to time-varying service distributions by Whitt \cite{whitt2018time}. In our implementation, arrivals and LOS vary daily, forming a discrete-time approximation of a non-stationary infinite-server queue. Eq. \ref{eq:1} holds under the assumptions that (i) arrivals follow a nonhomogeneous Poisson process, (ii) a patient's LOS is independent from other patients' LOS and from the arrival process, and (iii) the time-varying LOS distributions have finite mean at each point in time, so that the expected occupancy remains finite. These three assumptions are satisfied in our setting, allowing \( \rho_t \) to be computed as a convolution of past arrivals with the corresponding survival functions \citep{whitt2018time, fralix2010new}.

The assumption of a nonhomogeneous Poisson arrival process is supported by several diagnostics applied to site-level admission data. Kolmogorov–Smirnov tests of interarrival times rejected exponentiality at all sites, indicating that arrival rates are not constant over time. We then evaluated daily counts using the dispersion index, which compares the variance to the mean. Dispersion ratios were below but close to 1, and chi-squared p-values were near 1, showing no significant deviation from the variance structure expected under a Poisson process. Thus, despite time-inhomogeneity, marginal daily variability remains consistent with Poisson-like behavior. A chi-squared goodness-of-fit test rejected a homogeneous Poisson distribution at all sites, likely due to seasonality and structural patterns.

We therefore adopt a nonhomogeneous Poisson process with time-varying rate $\lambda_t$ within the infinite-server framework considered. The discrete approximation in Eq. \ref{eq:1} replaces the continuous convolution with a summation up to $S_{\max}$, defined for each site as the 99th percentile of the fitted LOS distribution. The tail probability $\mathbb{P}(S > u \mid \tau = t - u)$ is computed using the distribution-specific formulas in Section~\ref{LOSmodeling}, parameterized by smoothed $\mu_{t - u}$ and $\sigma^2_{t - u}$. Iterating over $t$ and $u \in \{0, \dots, S_{\max}\}$ yields the time series of expected occupancy $\rho_t$, which we use to evaluate capacity planning strategies.

To assess ICU bed requirements under uncertainty, we explore two planning strategies informed by $\rho_t$:

\begin{itemize}
    \item \textbf{Average Occupancy Estimation:} A baseline benchmark defined as
    \[
    B_{\text{average}} = \bar{\rho} + \sqrt{\bar{\rho}},
    \]
    where \( \bar{\rho} = \bar{\lambda} \cdot \mathbb{E}[S] \) is the product of the average admission rate \( \bar{\lambda} \) and the mean LOS \( \mathbb{E}[S] \), both computed over the full historical dataset. This approach does not rely on STL smoothing or time-series decomposition.

    The square-root term provides a buffer for random occupancy fluctuations, reflecting that under a Poisson process the standard deviation is approximately the square root of the mean. Similar buffers are common in operations research, including call center staffing \citep{borst2004dimensioning, koole2002queueing} (e.g., the square-root staffing rule in the Erlang-C model \citep{janssen2011refining, whitt2007you}) and public transit planning \citep{schmidt2024planning}. However, this method does not formally control the probability of exceeding capacity.

    \item \textbf{Overflow-Constrained Occupancy:} We formulate capacity planning as a constrained optimization problem in which we minimize the number of beds subject to a probabilistic performance constraint. Specifically, we solve:
    
\[
\min B \quad \text{s.t.} \quad \mathbb{P}(L_t > \gamma B) \leq \alpha,
\]

  where \( L_t \) denotes the number of occupied beds at time \( t \), \( \gamma \in (0,1) \) is a utilization threshold (e.g., 0.85), and \( \alpha \in (0,1) \) is the maximum acceptable overflow risk. Minimizing \( B \) promotes high utilization, while the constraint limits how often occupancy exceeds \( \gamma B \). This can be evaluated over the time series $\{\rho_t\}$ to measure how frequently a capacity level is exceeded over the planning horizon.

    Under the $M_t/G_t/\infty$ assumptions (NHPP arrivals and independent LOS), the number in system at time $t$ is Poisson with mean $\rho_t$, given by the convolution in Eq.~\ref{eq:1} \citep{whitt2018time}. Overflow probabilities are therefore computed using the Poisson tail:
    
   \begin{equation}
    \mathbb{P}(L_t > \gamma B) 
    = 1 - F_{\text{Poisson}}(\gamma B; \rho_t),
    \label{eq:2aa}
\end{equation}

    where $ F_{\text{Poisson}}(\gamma B; \rho_t) $ denotes the CDF of a Poisson distribution with mean \( \rho_t \) evaluated at $\gamma B$.

    This method also uses STL-smoothed arrival rates and LOS parameters. Unlike the other two, it directly constrains expected overflow risk and ensures that the probability of exceeding \( \gamma B \) remains below the acceptable limit \( \alpha \). It thus provides a capacity level with explicit resilience guarantees.    
\end{itemize}

\subsection{Future Projection of ICU Bed Requirements} \label{FutureProjection}

While retrospective NICU occupancy estimates inform historical demand and capacity alignment, planning requires a forward-looking perspective. To extend the framework in Section~\ref{OccupancyEstimation}, we develop a births-driven projection method combining regional birth forecasts with historical site-level admission shares and seasonal arrival and LOS patterns. This helps ensure that long-term capacity planning reflects demographic trends and observed temporal variability.

Let $Y_f = \{ y_{\min}, \ldots, y_{\max} \}$ denote the desired projection years. We use two historical windows:  
(i) a set $Y_h^\omega$ of recent years to determine baseline admissions and site shares, and  
(ii) a set $Y_h^\nu$ of historical years to provide reference patterns for within-year arrivals and LOS.  
The first captures recent structural conditions, while the second preserves realistic daily variability.

Annual demand is defined relative to historical admissions and scaled to projected births. Let $\tilde{K}_y$ denote projected total births in year $y \in Y_f$, obtained from external forecasts. Baseline system-wide admissions are the mean over $Y_h^\omega$,
\[
A_{\text{base}} = \frac{1}{|Y_h^\omega|} \sum_{y \in Y_h^\omega} A_y,
\]
where $A_y$ is observed admissions in year $y$. Let $y_0 = y_{\min}$ denote the base year. Projected admissions for $y \in Y_f$ are given by

\begin{equation}
\hat{A}_y = A_{\text{base}} \cdot \left( \frac{\tilde{K}_y}{\tilde{K}_{y_0}} \right)^{\eta} \cdot \psi^{(y-y_0)}.
\label{eq:3}
\end{equation}

Here, $\eta$ denotes the elasticity of admissions with respect to births ($\eta = 1$ implies proportional scaling), and $\psi$ is a structural drift factor capturing multiplicative annual changes. This formulation is used for two reasons. First, it reflects the dependence of admissions on birth volumes, with $\eta$ governing responsiveness (e.g., proportional change when $\eta = 1$, slower growth when $\eta < 1$). Second, $\psi$ captures long-term shifts not explained by births, such as changes in admission practices or referral patterns. Together, these parameters provide a transparent model that accounts for demographic and structural drivers of long-run demand.

To allocate $\hat{A}_y$ across sites, we compute the average site share over $Y_h^\omega$:
\[
s_m = \frac{1}{|Y_h^\omega|} \sum_{y \in Y_h^\omega} \frac{A_{m,y}}{\sum_{m'} A_{m',y}}, 
\quad \sum_m s_m = 1,
\]
where $A_{m,y}$ is admissions at site $m$ in year $y \in Y_h^\omega$. Projected admissions at site $m$ in year $y \in Y_f$ are then $\hat{A}_{m,y} = s_m \cdot \hat{A}_y$.

For each site $m$ and projection year $y \in Y_f$, we preserve intra-annual variation by resampling patterns from $Y_h^\nu$. A reference year $h$ is sampled at random, and its daily admissions $\lambda_{m,h,t}$ are normalized to construct a profile
\[
p_{m,h,t} = 
\begin{cases}
\frac{\lambda_{m,h,t}}{\sum_{u=1}^{365} \lambda_{m,h,u}}, & \text{if } \sum_{u} \lambda_{m,h,u} > 0, \\
\frac{1}{365}, & \text{if } \sum_{u} \lambda_{m,h,u} = 0,
\end{cases}
\quad t = 1,\ldots,365.
\]
We define the projected daily arrivals $\hat{\lambda}_{m,y,t}$ by scaling the normalized profile $p_{m,h,t}$ so that its sum matches the annual admission target $\hat{A}_{m,y}$. Therefore, 
\[
\hat{\lambda}_{m,y,t} = \hat{A}_{m,y} \cdot p_{m,h,t}, 
\quad \sum_{t=1}^{365} \hat{\lambda}_{m,y,t} = \hat{A}_{m,y}.
\]

To project LOS, we resample a reference year $h' \in Y_h^\nu$ and carry forward its daily mean $\mu_{m,h',t}$ and variance $\sigma^2_{m,h',t}$ without scaling, preserving intra-annual variation. The site-specific LOS family identified in Section~\ref{LOSmodeling} is assumed fixed over the horizon, with parameters determined from the resampled mean and variance series. Our data indicate that the LOS distribution is stable across years, and this method retains that stability while preserving seasonal variation.

Given projected daily arrivals $\hat{\lambda}_{m,y,t}$ and the LOS distribution, site-specific expected occupancy at site $m$ in year $y$ on day $t$ is estimated using the $M_t/G_t/\infty$ convolution:
\[
\hat{\rho}_{m,y,t} = \sum_{u=0}^{S_{\max}} \hat{\lambda}_{m,y,t-u} \cdot {\mathbb{P}}(\hat{S} > u \mid \tau = t-u),
\]
where $\hat{S}$ is the estimated LOS and $S_{\max}$ is the site-specific truncation horizon from the fitted LOS model in Section~\ref{LOSmodeling}. The conditional survival probability ${\mathbb{P}}(\hat{S} > u \mid \tau = t-u)$ is computed from the site's best-fitting parametric LOS distribution, parameterized by the resampled reference year.

Our framework links empirical demand estimation with analytical queueing approximations and provides a flexible, interpretable methodology for ICU planning. Each component corresponds to observable quantities such as admission rates and LOS distributions, and occupancy follows from a transparent convolution formula. Unlike black-box models, assumptions can be examined independently, and capacity thresholds are defined through explicit probabilistic constraints. Such transparency is emphasized in healthcare operations research for validation and policy adoption \citep[e.g.,][]{gans2003telephone, koole2023practice, hyndman2018forecasting}.

The framework also accommodates site-level heterogeneity and supports scenario-based evaluation under time-varying conditions. These scenarios allow decision makers to benchmark historical performance, consider worst-case behavior, and design capacity with resilience guarantees. It further extends to future years by combining demographic projections with historical site-specific arrival and LOS patterns, enabling long-term planning under plausible demand trajectories.

\section{Results} \label{Results}

In this section, we present the results of our modeling pipeline which consists of results from STL decomposition, LOS distribution modelling, and the evaluation of bed occupancy scenarios across the NICU sites.

\subsection{STL Grid Search and Smoothed Trend Estimation}

Table \ref{tab:stl_params} reports the optimal STL configuration selected via grid search. The same parameters are chosen across all five NICU sites. Shorter seasonal and trend windows (7 and 15 days) best capture the temporal dynamics, with the weekly seasonal window reflecting short-term cycles and the bi-weekly trend window balancing smoothness and responsiveness. Linear fitting (degree 1, standard LOESS) with non-robust weighting minimizes residual variance in both admission and LOS time series, suggesting that local trends dominate the time series structure. For LOS variance estimation, a 31-day rolling window applied to STL residuals produces the most stable standard deviation estimates. This wider window likely balances responsiveness to variance shifts while reducing noise from small daily LOS samples and smoothing transient outliers.

\begin{table}[h!]
\centering
\small
\caption{Optimal STL decomposition parameters selected across all NICU sites}
\begin{tabular}{lc}
\toprule
Parameter & Value \\
\midrule
Seasonal Window & 7 days  \\
Trend Window & 15 days  \\
Seasonal Degree & 1 \\
Trend Degree & 1 \\
Robust & False  \\
Rolling Window (LOS Variance) & 31  \\
\bottomrule
\end{tabular}
\label{tab:stl_params}
\end{table}

\begin{figure}[!ht]
\centering
\begin{subfigure}[t]{0.48\textwidth}
\centering
\includegraphics[width=\textwidth]{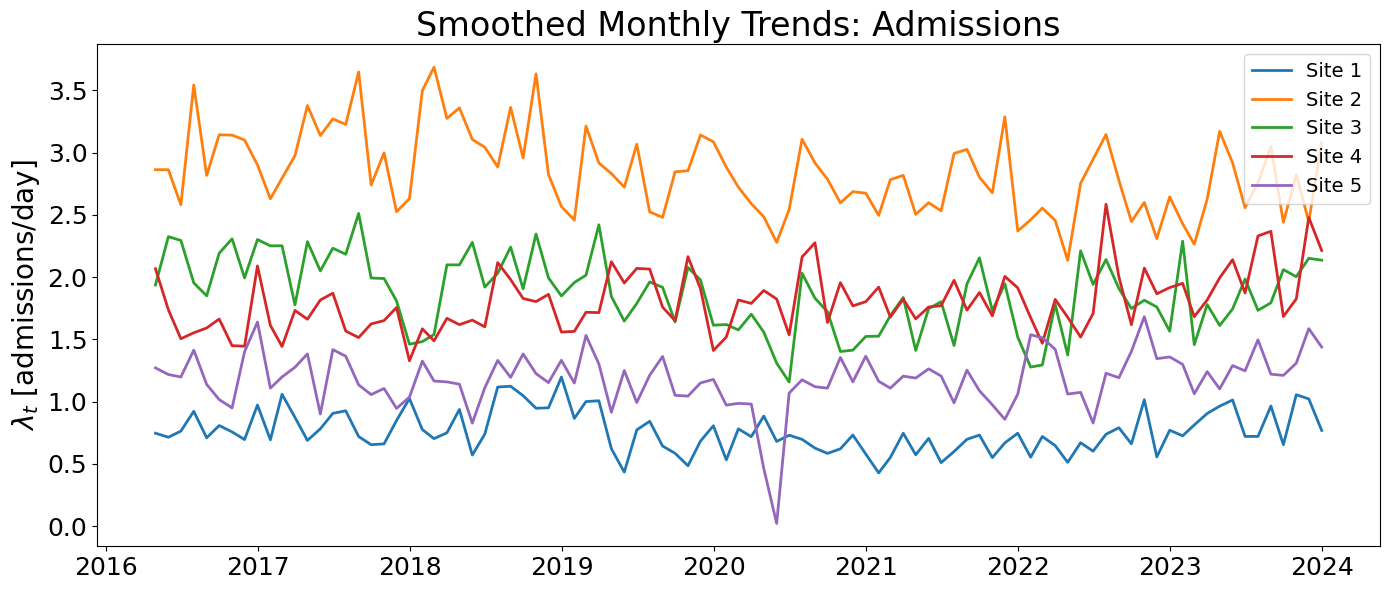}
\caption{Smoothed monthly admission rate $\lambda_t$}
\end{subfigure}
\hfill
\begin{subfigure}[t]{0.48\textwidth}
\centering
\includegraphics[width=\textwidth]{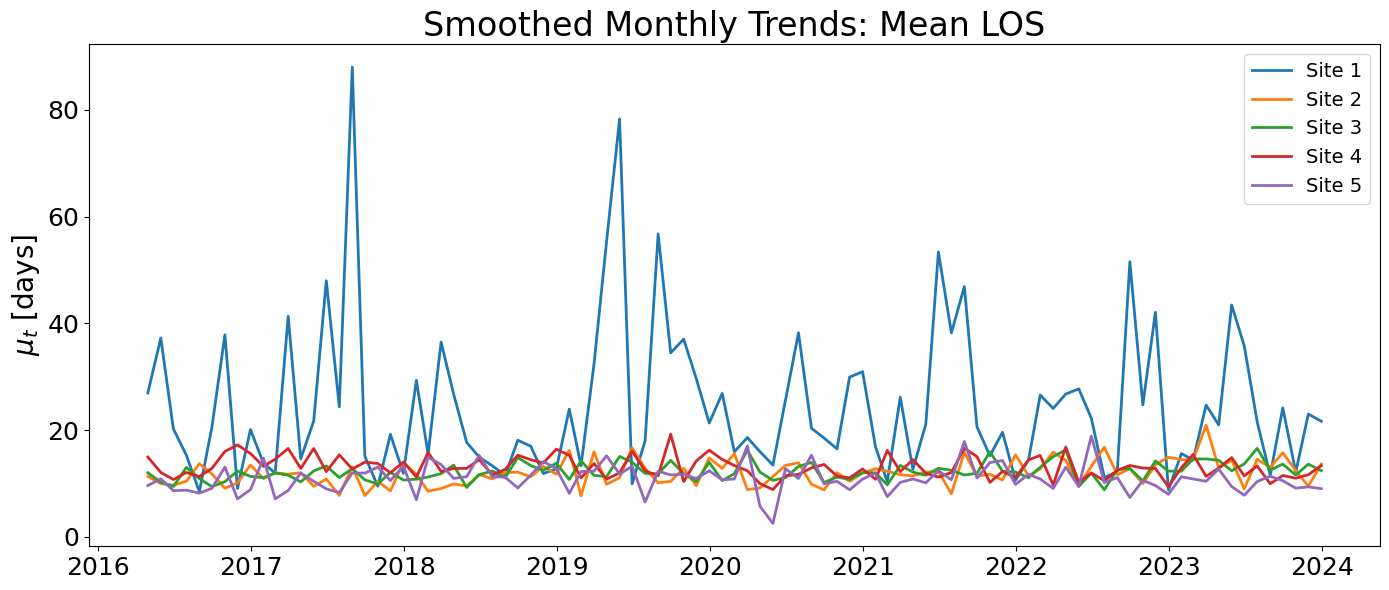}
\caption{Smoothed monthly mean LOS $\mu_t$}
\end{subfigure}

\caption{Smoothed monthly trends in admission rate and mean LOS across the five NICU sites}
\label{fig:monthly_trends}
\end{figure}

Figure \ref{fig:monthly_trends} shows smoothed monthly trends for admission rate $\lambda_t$ and mean LOS $\mu_t$ at each site. Admission rates display recurring variation and site-level differences, while mean LOS is relatively stable except at Site 1, which shows higher variability and occasional spikes. Daily STL outputs are aggregated to monthly values for visualization. However, all modeling steps are conducted using daily resolution data.

\subsection{Parametric LOS Distribution Fit and Comparison}

As discussed in Section \ref{LOSmodeling}, we fit five candidate parametric distributions to empirical LOS data at each site. Table \ref{tab:los_fit_summary} reports the selected distribution per site, its shape parameter $\kappa$ (if applicable), the RMSE relative to the Kaplan-Meier curve, and the truncation threshold $S_{\max}$. Site-specific selections reflect differences in tail behavior, with Fisk capturing heavier tails at Sites 1 and 2, and Exponential fitting intermediate decay at Site 5.

\begin{table}[!ht]
\centering
\caption{Best-fitting LOS distribution by site, along with RMSE and fixed shape parameter $\kappa$}
\label{tab:los_fit_summary}
\begin{tabular}{ccccc}
\toprule
Site & Best Distribution & RMSE & $\kappa$ & $S_{\max}$ \\
\midrule
1 & Fisk & 0.01 & 1.34 & 325 \\
2 & Fisk & 0.03 & 1.25 & 340 \\
3 & Exponential & 0.02 & -- & 40\\
4 & Gamma & 0.01 & 1.16 & 42\\
5 & Exponential & 0.02 & -- & 39\\
\bottomrule
\end{tabular}
\end{table}

\begin{figure}[!ht]
\centering

\begin{subfigure}[t]{0.6\textwidth}
\centering
\includegraphics[width=\textwidth]{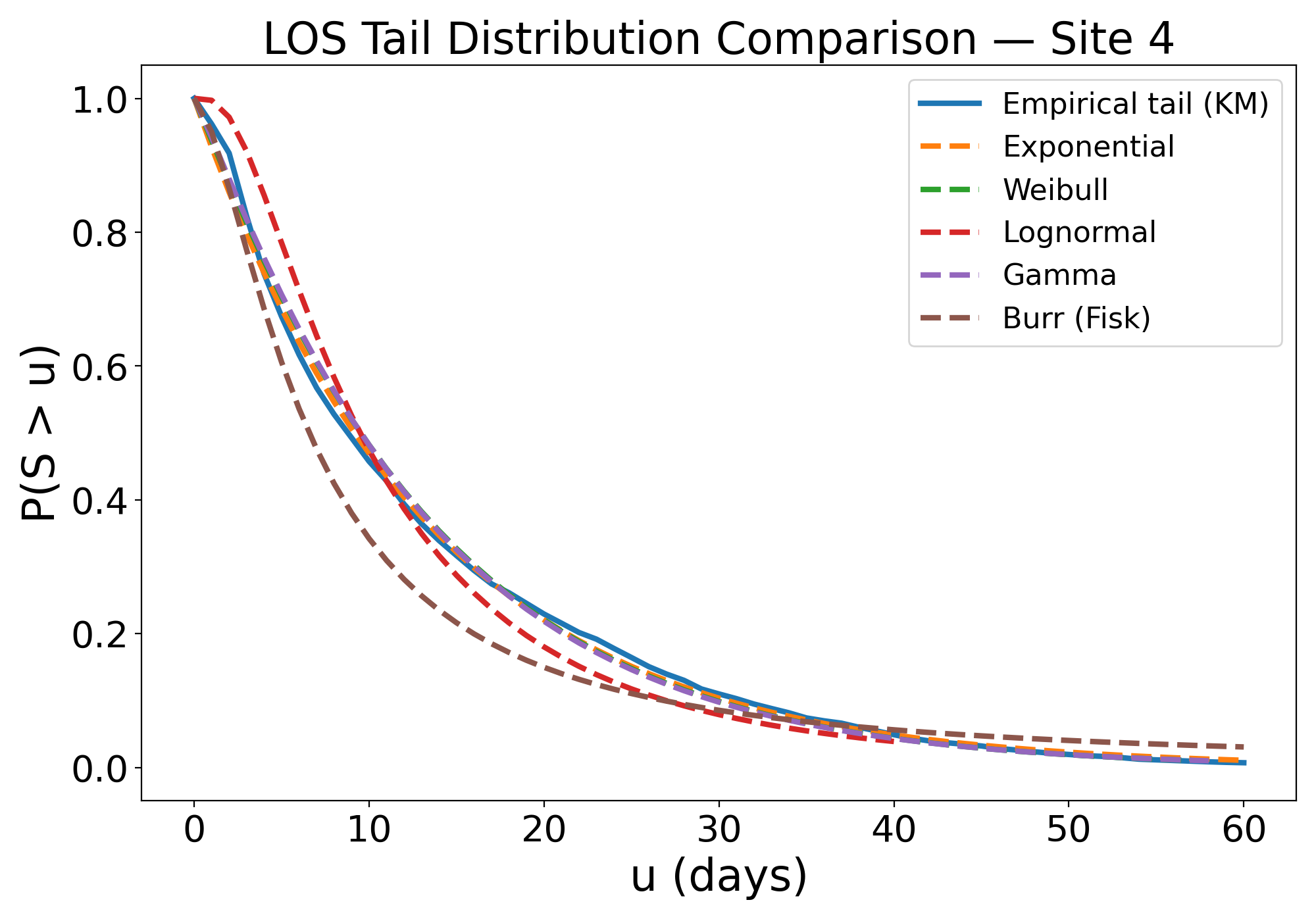}
\caption{Site 4}
\end{subfigure}

\caption{Comparison of empirical LOS tail probabilities $\mathbb{P}(S > u)$ with candidate parametric distributions at an example site}
\label{fig:tail_comparisons}
\end{figure}

Figure \ref{fig:tail_comparisons} compares Kaplan-Meier LOS tails with the five parametric distributions at a represntetaive site.  The plot shows the marginal tail probability $\mathbb{P}(S > u)$ up to 60 days, chosen for visualization purposes only, as over 90\% of admissions have LOS below this threshold. The selected distributions closely match empirical tails across most time points, which supports their use in our occupancy modeling. Similar patterns are observed across the other sites and are shown in Figure \ref{fig:tail_comparisons2}.

\subsection{Scenario-Based Capacity Estimates and Site-Level Utilization}

\subsubsection*{\textbf{Capacity Estimation Strategies}}

We evaluate the traditional average occupancy estimate and the overflow-constrained capacity strategies given in Section~\ref{OccupancyEstimation}. For the latter, we compute \(B_{0.01}\) and \(B_{0.05}\), the minimum number of beds ensuring daily exceedance probabilities of at most 1\% and 5\%, respectively. These strategies give insights into trade-offs between resilience and utilization, with smaller \( \alpha \) yielding more conservative capacity. We set \( \gamma = 1 \) in Eq.~\ref{eq:2aa}, targeting overflow relative to full nominal bed capacity \( B \). This parameter can be adjusted. For example, \( \gamma = 0.85 \) evaluates risk of exceeding 85\% capacity, consistent with common operational guidelines \citep{janke2022hospital}.

Table~\ref{tab:bed_estimates_vs_actual} compares the actual number of beds at each NICU site with proposed capacity estimates. The traditional estimate \(B_{\text{average}}\) yields the smallest values and later is shown to underestimate capacity needed for resiliency under demand variability. These comparisons use the adjusted demand series from Section~\ref{ProblemData}, which remove capacity-driven transfers from Site 2. Consequently, the actual number of beds at Site~2 appears lower than required, while Sites~3–5 appear higher relative to adjusted demand, reflecting historical redistribution within the network.

\begin{table}[!ht]
\centering
\caption{Comparison of actual beds with scenario-based estimates}
\label{tab:bed_estimates_vs_actual}
\begin{tabular}{lcccc}
\toprule
Site  & Actual Beds & \( B_{\text{average}} \) & \(B_{0.05}\) & \(B_{0.01}\)  \\
\midrule
1 & 14 & 20 & 26 & 31  \\
2 & 39 & 67 & 83 & 90  \\
 3 & 30 & 14 & 20 & 24  \\
4 & 27 & 16 & 22 & 26  \\
 5 & 16 & 11 & 15 & 19  \\
\bottomrule
\end{tabular}
\end{table}

\subsubsection*{\textbf{Utilization Outcomes across Sites}}

We next examine how bed counts from each strategy affect site-level utilization. Table~\ref{tab:utilization_comparison} reports the mean and standard deviation of daily utilization over the entire time horizon of our dataset, assuming each proposed capacity is implemented. We also report weighted mean and standard deviation of daily utilization using admission volume as weights, so higher-volume sites contribute proportionally to system-wide performance.

\begin{table}[!ht]
\captionsetup{font=small}
\caption{Utilization under different capacity planning strategies (mean and std)}
\centering
\footnotesize
\bgroup
\def\arraystretch{0.2}
\begin{tabular}{p{1.5cm}p{1.87cm}p{2cm}p{2cm}p{2cm}}
\toprule
Site & \( B_{\text{average}} \) & \(B_{0.05}\) & \(B_{0.01}\)  \\
\midrule
 1 &  82.45 (14.38) & 63.43 (11.06) & 53.20 (9.28)  \\
 2 &  91.10 (12.86) & 73.54 (10.38) & 67.82 (9.57)  \\
 3 & 83.00 (27.85) & 58.10 (19.50) & 48.42 (16.25)  \\
 4 &  82.10 (22.88) & 59.71 (16.64) & 50.52 (14.08)  \\
 5 & 76.47 (27.76) & 56.08 (20.36) & 44.27 (16.07)  \\
\midrule
Weighted site-level &  85.07 (5.24) & 64.74 (7.31) & 56.36 (9.50)  \\
\bottomrule
\end{tabular}
\egroup
\label{tab:utilization_comparison}
\end{table}

We observe that the traditional heuristic maintains high utilization but offers limited surge protection. In contrast, overflow-constrained strategies (\( B_{0.05} \) and \( B_{0.01} \)) add beds to limit daily exceedance probabilities to 5\% and 1\%, reducing mean utilization while accounting for demand fluctuations. At the system level, weighted utilization confirms these patterns. The traditional estimate yields average utilization of 85.07\% with low variance, which has close alignment with current bed levels. More conservative strategies reduce mean utilization, from 64.74\% under \( B_{0.05} \) to 56.36\% under \( B_{0.01} \). These results illustrate the tradeoff between operational efficiency and capacity resilience.

\begin{figure}[!ht]
    \centering

    \begin{subfigure}[t]{0.49\textwidth}
        \centering
        \includegraphics[width=\textwidth]{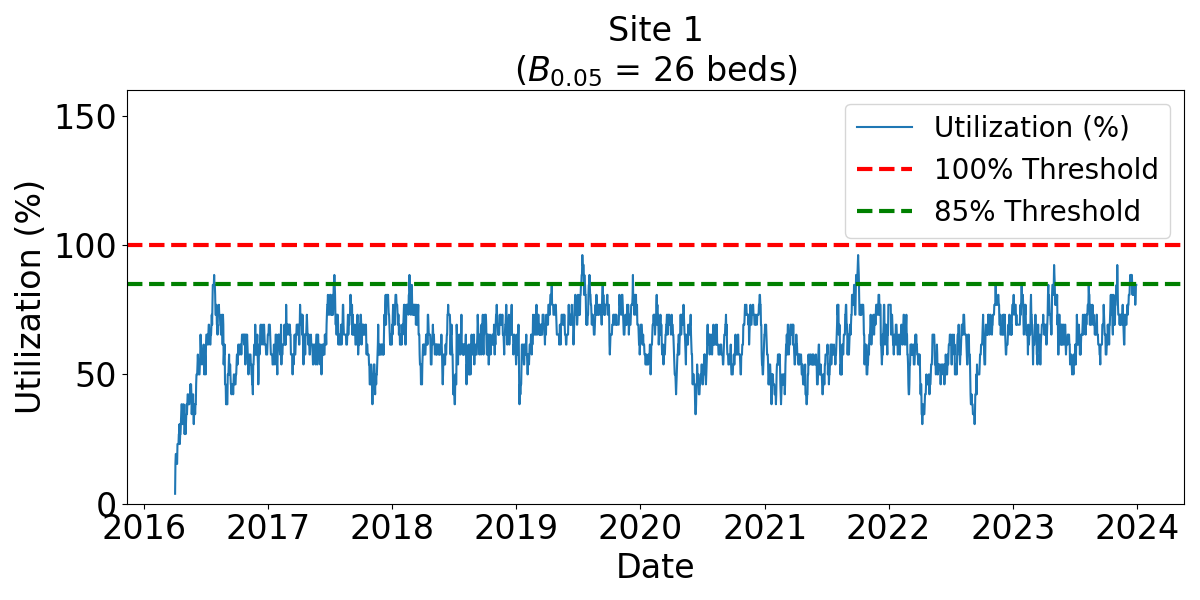}
        \caption{Site 1}
    \end{subfigure}
    \hfill
    \begin{subfigure}[t]{0.49\textwidth}
        \centering
        \includegraphics[width=\textwidth]{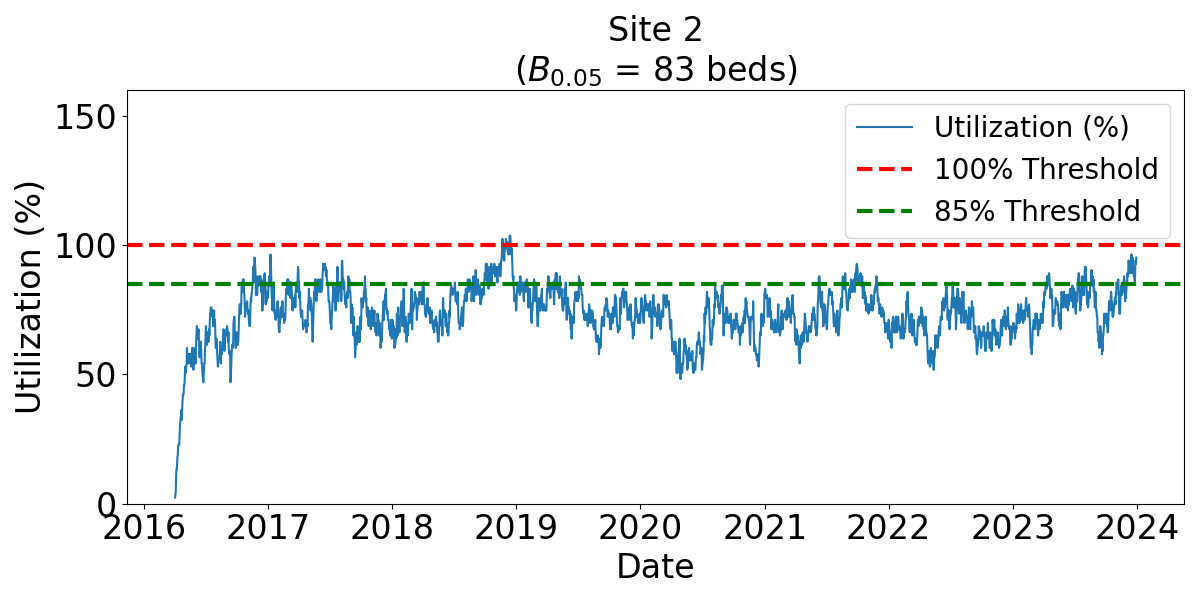}
        \caption{Site 2}
    \end{subfigure}
% \\[0.5em]
    \begin{subfigure}[t]{0.49\textwidth}
        \centering
        \includegraphics[width=\textwidth]{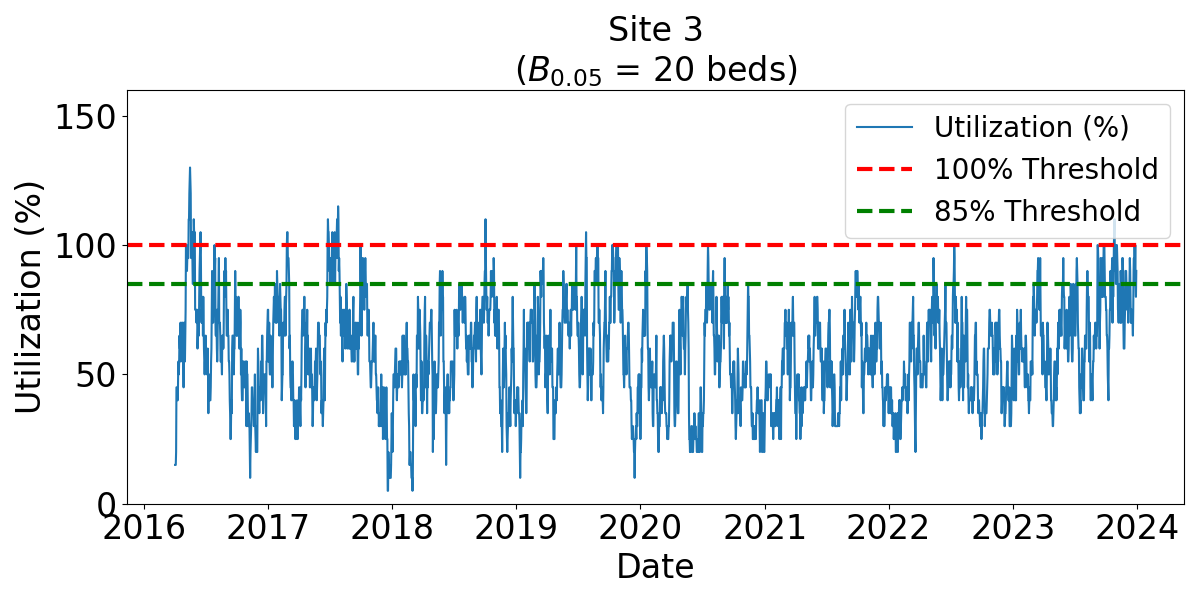}
        \caption{Site 3}
    \end{subfigure}
    \hfill
    \begin{subfigure}[t]{0.49\textwidth}
        \centering
        \includegraphics[width=\textwidth]{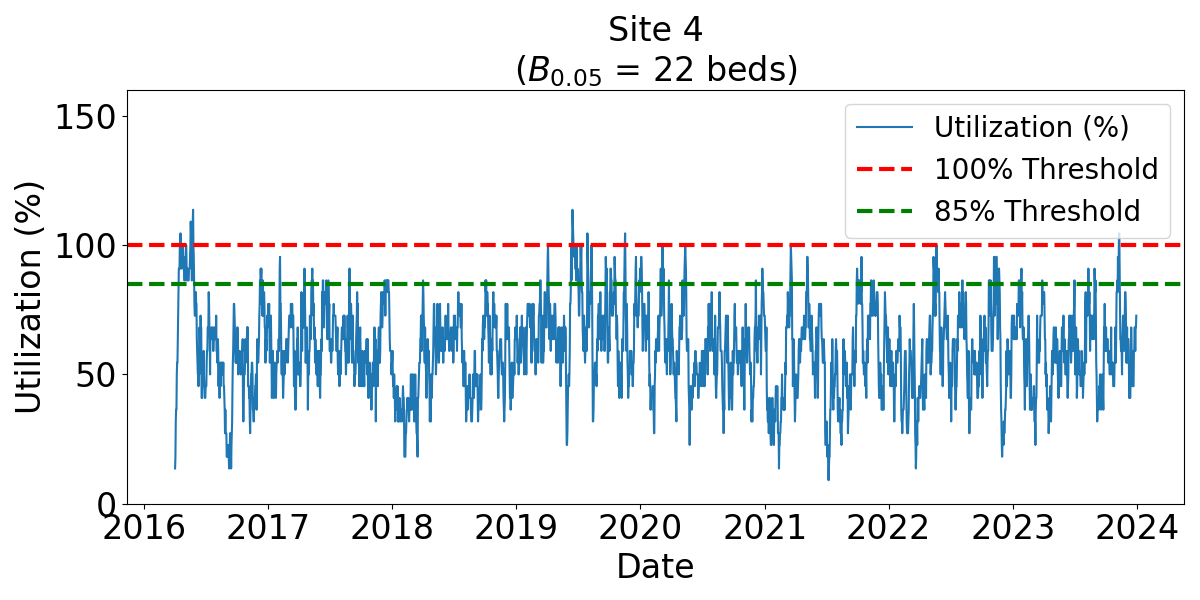}
        \caption{Site 4}
    \end{subfigure}
% \\[0.5em]
    \begin{subfigure}[t]{0.49\textwidth}
        \centering
        \includegraphics[width=\textwidth]{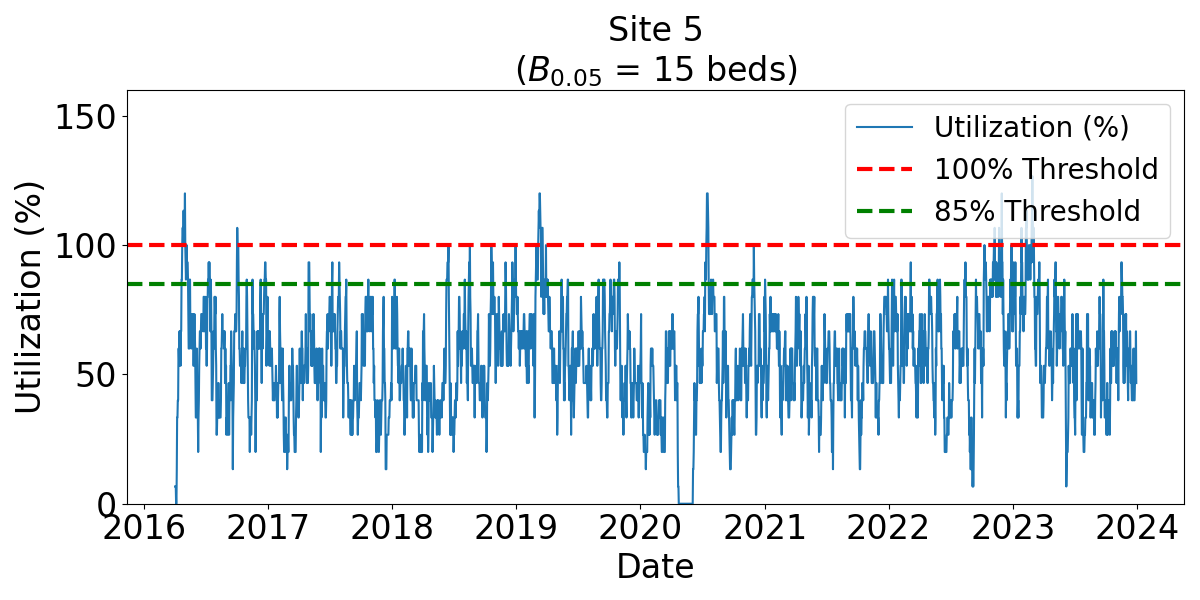}
        \caption{Site 5}
    \end{subfigure}
% \\[0.5em]
    \caption{Daily bed utilization (\%) under overflow-constrained threshold \(B_{0.05}\). Dashed lines show 85\% and 100\% thresholds.}
    \label{fig:utilization_b05}
\end{figure}

Figure~\ref{fig:utilization_b05} shows site-level utilization under the \(B_{0.05}\) scenario. Sites with greater demand variability exhibit some exceedances beyond 85\% utilization, even with added capacity, but most days remain within safe margins. The observed upper bound, typically near 85\% across institutions, suggests these estimates are appropriate for planning.

\begin{figure}[!ht]
\centering
    \begin{subfigure}[t]{0.49\textwidth}
        \centering
        \includegraphics[width=\textwidth]{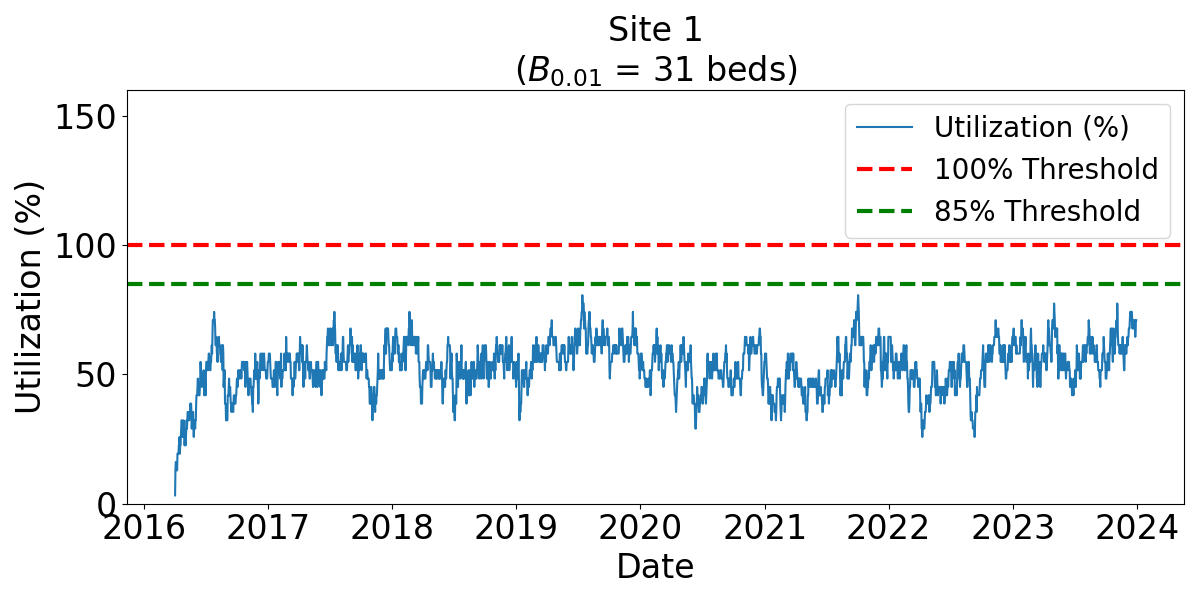}
        \caption{Site 1}
    \end{subfigure}
    \hfill
    \begin{subfigure}[t]{0.49\textwidth}
        \centering
        \includegraphics[width=\textwidth]{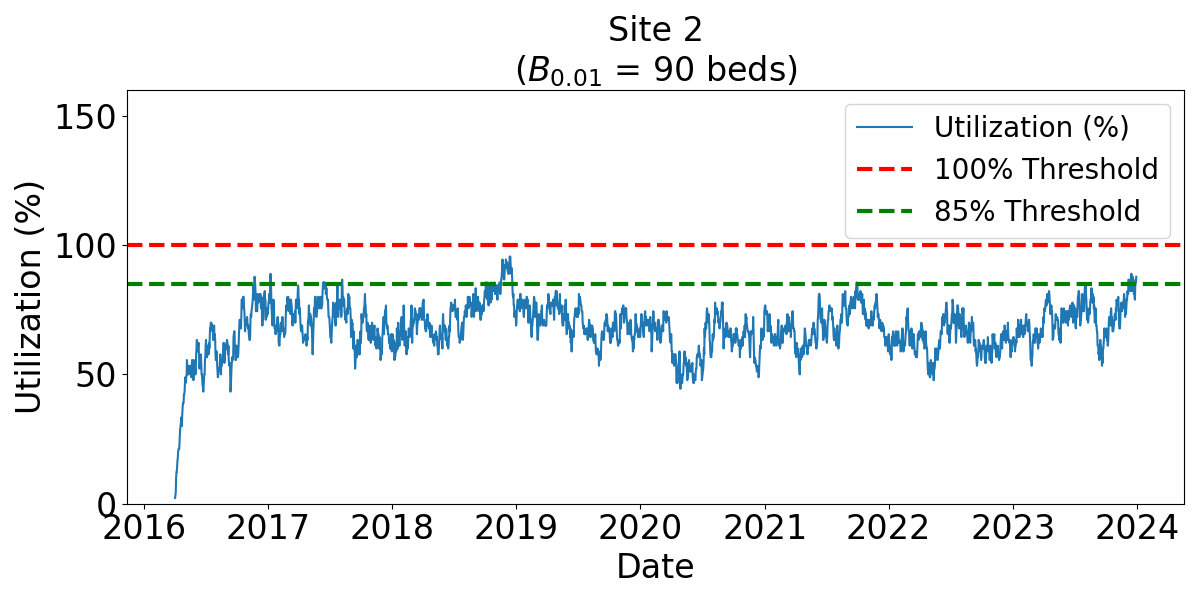}
        \caption{Site 2}
    \end{subfigure}
% \\[0.5em]
    \begin{subfigure}[t]{0.49\textwidth}
        \centering
        \includegraphics[width=\textwidth]{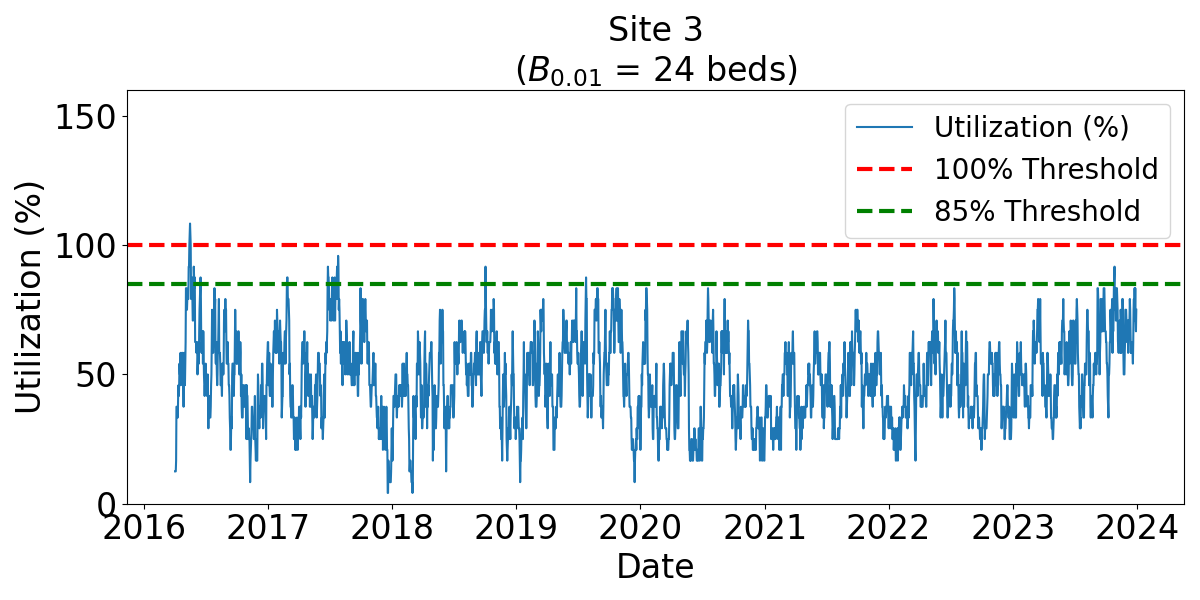}
        \caption{Site 3}
    \end{subfigure}
    \hfill
    \begin{subfigure}[t]{0.49\textwidth}
        \centering
        \includegraphics[width=\textwidth]{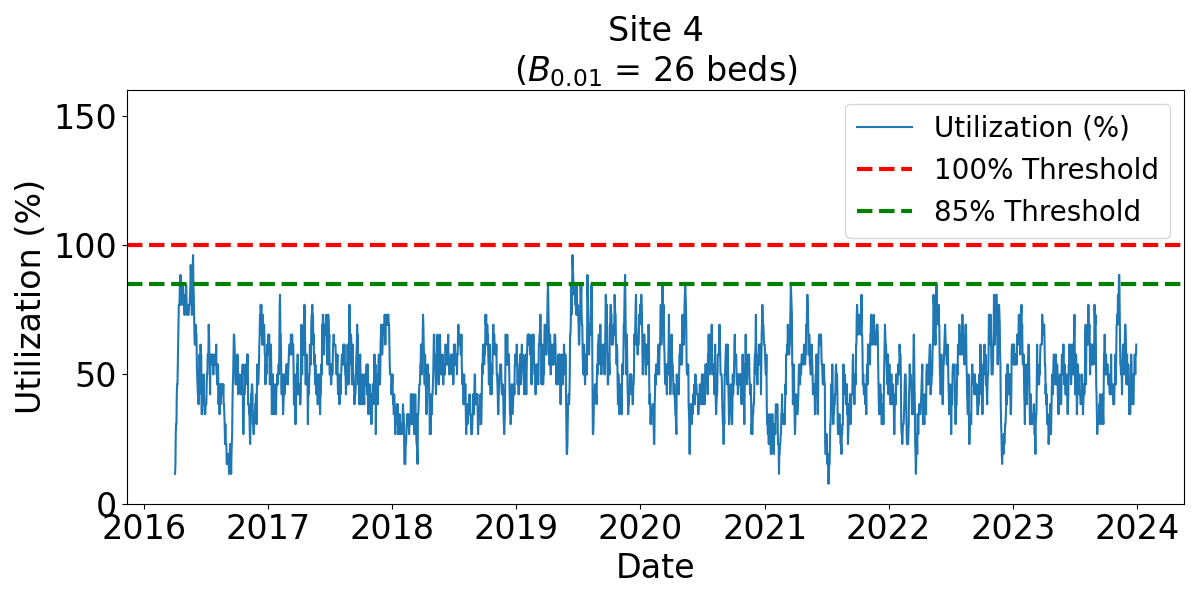}
        \caption{Site 4}
    \end{subfigure}
% \\[0.5em]
    \begin{subfigure}[t]{0.49\textwidth}
        \centering
        \includegraphics[width=\textwidth]{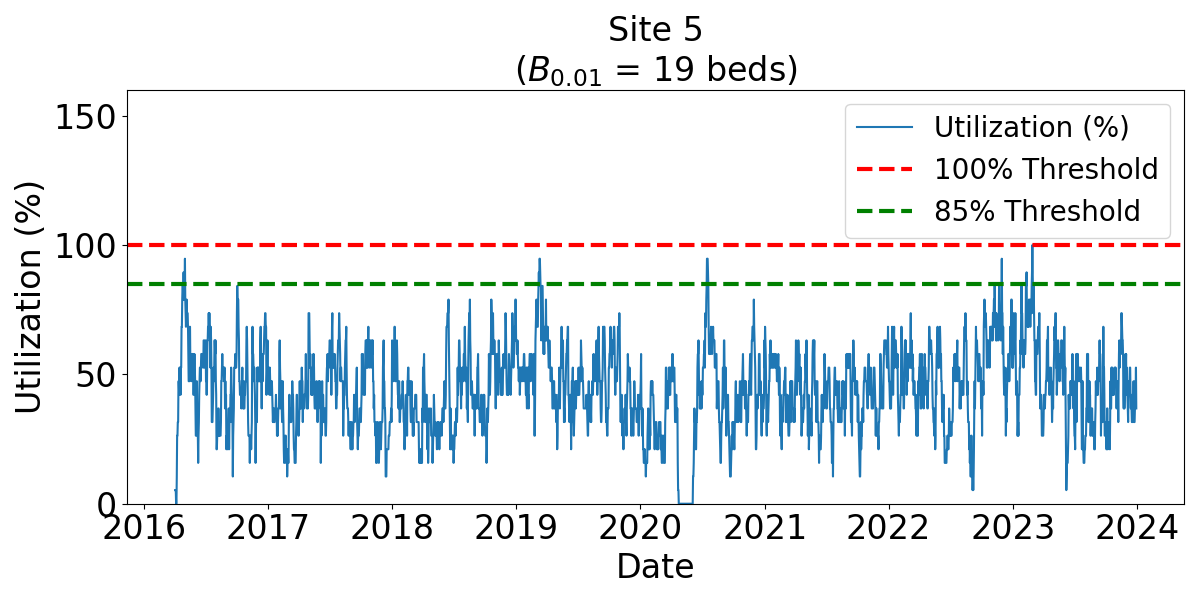}
        \caption{Site 5}
    \end{subfigure}
\caption{Daily utilization under the $B_{0.01}$ planning strategy. Red and green lines indicate 100\% and 85\% occupancy, respectively.}
\label{fig:site_utilization_1p}
\end{figure}

We also examine utilization patterns under $B_{0.01}$. Figure~\ref{fig:site_utilization_1p} shows daily utilization trajectories across the five NICU sites. The \( B_{0.01} \) scenario provides stronger surge protection and substantially lowers utilization, often below 70\%. While it reduces overflow risk, it may lead to persistent underutilization during typical periods, which reflects its more conservative capacity buffer. The utilization threshold parameter \( \gamma \) in Eq.~\ref{eq:2aa} offers additional flexibility for planners. Allowing \( \gamma > 1 \) permits temporary surges beyond nominal bed capacity within the resiliency constraint, which can increase average utilization while tolerating controlled exceedances. Such a specification may be relevant in settings where short-term overflow can be absorbed operationally, for example through temporary surge beds or flexible staffing arrangements, thereby providing an intermediate option between strict capacity limits and purely utilization-based planning.

A key insight from our modeling pipeline is the influence of time-varying LOS variance on occupancy. Many models rely only on mean LOS, but we estimate both mean and variance using STL decomposition and rolling residuals. Variance is incorporated into the $M_t/G_t/\infty$ model when the lognormal distribution is selected, since its parameterization depends on both moments. This allows the framework to capture both the expected load and its dispersion over time and can help yielding more realistic capacity estimates. The modular structure also allows evaluation of alternative resilience targets while remaining aligned with historical utilization patterns.

As discussed in Section~\ref{LitReview}, static heuristics such as the 85\% rule have limitations for ICU planning. Although this rule targets 85\% average utilization, it assumes stationary and evenly distributed demand over time. In practice, however, demand varies substantially day to day. At Site 3, which exhibits high variability in daily demand, occupancy exceeded 100\% of planned capacity on over 23\% of days, despite using the 85\% rule, which suggests 14 beds. Our empirical analyses reveal that this steady-state estimate fails to accommodate temporal surges, leading to frequent overload. At the same time, 31.68\% of days fell below 70\% occupancy, with an average shortfall of 18.48\% (with a standard deviation of 11.95\%), indicating significant under-utilization.

In contrast, the least conservative strategy, \( B_{0.05} \), identifies 20 beds for Site 3 (see Table~\ref{tab:bed_estimates_vs_actual}) and limits overcapacity risk to 5\%. Under this threshold, average utilization falls to 58.10\%, and exceedances above 100\% become rare. These results illustrate the trade-off between high long-term utilization and protection against temporal overload, as lower overload risk requires additional capacity and reduces average utilization.

Therefore, although the 85\% rule may serve as a rule of thumb, our findings support cautions raised by Bain et al.\ \cite{bain2010myths} and Au et al.\ \cite{au2009predicting} that applying it without accounting for temporal variation can underestimate capacity needs. Increasing beds reduces overload frequency but lowers average utilization, which highlights the trade-off between efficiency and resilience.

\subsubsection*{\textbf{Forward-looking Scenarios and Projections}}

In addition to retrospectively analyzing utilization and capacity mismatches, our framework supports forward-looking scenario planning. Using Eq. \ref{eq:1}, we evaluate alternative assumptions by applying a multiplier \(\beta\) to \(\lambda_t\), \(\mu_t\), or \(\sigma_t^2\). For example, seasonal surges can be modeled by increasing \(\lambda_t\), staffing shortages that delay discharges  by increasing \(\mu_t\), and improved policies and practices by reducing \(\mu_t\).

Although changes in mean admissions and LOS have intuitive capacity effects, the role of LOS variance is less visible yet important for affecting the right tail of occupancy distributions, especially in high-utilization environments such as NICUs. We conduct a sensitivity analysis by scaling the variance of the fitted lognormal LOS (as this distribution uses both moments) using $\beta$, while holding the mean fixed. We consider \(\beta \in \{0.2, 0.5, 0.8\}\) for reduced variability, \(\beta \in \{1.2, 1.5, 1.8\}\) for increased variability, and \(\beta = 0\) as a zero-variance setting in which all patients have identical LOS durations. For each site, required beds are re-estimated under each $\beta$ and compared to the baseline $\beta=1$. Table~\ref{tab:variance_sensitivity} reports percentage changes relative to this baseline.

\begin{table}[!ht]
\centering
\caption{Percentage change in number of beds required under varying LOS variance multipliers \(\beta\), relative to the baseline case \(\beta = 1\). All scenarios use lognormal LOS distributions with fixed mean and scaled variance.}
\small
\def\arraystretch{0.7}
\begin{tabular}{clrrrrrrr}
\toprule
Site & Strategy & $\beta{=}0$ & $\beta{=}0.2$ & $\beta{=}0.5$ & $\beta{=}0.8$ & $\beta{=}1.2$ & $\beta{=}1.5$ & $\beta{=}1.8$ \\
\midrule
\multirow{2}{*}{1} & \(B_{0.05}\)        & 10.34 & 6.90 & 3.45 & 0    & -3.45 & -3.45 & -6.90 \\
                   & \(B_{0.01}\)        & 14.71 & 8.82 & 5.88 & 2.94 & 0     & -2.94 & -5.88 \\ 
\addlinespace[2pt]
\cmidrule{1-9}
\addlinespace[2pt]
\multirow{2}{*}{2} & \(B_{0.05}\)        & 10.98 & 6.10 & 3.66 & 1.22 & 0     & -1.22 & -2.44 \\
                   & \(B_{0.01}\)        & 15.38 & 7.69 & 3.30 & 1.10 & -1.10 & -2.20 & -3.30 \\
\addlinespace[2pt]
\cmidrule{1-9}
\addlinespace[2pt]
\multirow{2}{*}{3} & \(B_{0.05}\)        & 9.52  & 4.76 & 4.76 & 0    & 0     & 0     & 0 \\
                   & \(B_{0.01}\)        & 7.69  & 3.85 & 0    & 0    & 0     & -3.85 & -3.85 \\
\addlinespace[2pt]
\cmidrule{1-9}
\addlinespace[2pt]
\multirow{2}{*}{4} & \(B_{0.05}\)        & 4.35  & 4.35 & 0    & 0    & 0     & -4.35 & -4.35 \\
                   & \(B_{0.01}\)        & 11.11 & 7.41 & 3.70 & 3.70 & 0     & 0     & 0 \\
\addlinespace[2pt]
\cmidrule{1-9}
\addlinespace[2pt]
\multirow{2}{*}{5} & \(B_{0.05}\)        & 6.25  & 6.25 & 6.25 & 0    & 0     & 0     & 0 \\
                   & \(B_{0.01}\)        & 10.00 & 5.00 & 5.00 & 5.00 & 0     & 0     & 0 \\
\bottomrule
\end{tabular}
\label{tab:variance_sensitivity}
\end{table}

Table~\ref{tab:variance_sensitivity} shows a consistent pattern across sites and thresholds. Reducing LOS variance (\(\beta < 1\)) increases beds required to meet a fixed overflow target, whereas increasing variance (\(\beta > 1\)) decreases capacity needs. This contrasts with standard queueing intuition, where greater service duration variability is often linked to worse congestion \citep{whitt1983queueing}. For example, under \(B_{0.05}\) at Site 1, deterministic LOS (\(\beta=0\)) increases required capacity by 10.34\%, while inflating variance by 80\% reduces it by 6.90\%. Similar directional effects appear across the remaining sites and under \(B_{0.01}\).

This behavior follows from the structure of the \(M_t/G_t/\infty\) model. With mean LOS fixed, higher variance disperses discharges and reduces synchronized overlap among patient cohorts, thereby lowering peak occupancy. Lower variance concentrates discharges and increases temporal overlap, which raises capacity needed to meet the same overflow constraint. Although variability reduction is often viewed as beneficial \citep{hopp2011factory}, our results show that, in an infinite-server setting, LOS dispersion can affect risk-based capacity estimates.

We conclude this section by presenting projected bed requirements based on the approach in Section~\ref{FutureProjection}. We generate $R=300$ scenarios, each independently resampling historical years $h,h' \in Y_h^\nu$ for arrivals and LOS, while preserving births-driven annual admissions. For each site and year, we compute required capacity under $B_{\text{average}}$, $B_{0.05}$, and $B_{0.01}$.

We report two projection exercises: (a) retrospective validation and (b) forward-looking projection. Both use the same methodology but differ in reference and projection windows. For retrospective validation, $Y_h^\nu$ covers 2017–2021, and for forward projection, it covers 2017–2023 (data for 2016 is excluded as a reference year because the first three months were unavailable). In each case, $Y_h^\omega$ consists of the three most recent years within $Y_h^\nu$ to reflect current structural conditions in admissions and site shares. The window $Y_h^\nu$ offers a broader set of reference patterns for within-year arrivals and LOS and thus preserves realistic seasonal and intra-annual variability consistent with observed historical behavior. For retrospective validation, we set $y_{\min}=2022$ and $y_{\max}=2023$, and for forward projection, we set $y_{\min}=2024$ and $y_{\max}=2030$.

The projected total births $\tilde{K}_y$ are obtained from the Alberta Interactive Health Data repository \citep{albertadata} for the Calgary Zone. According to their documentation, birth projections use (i) female population forecasts and (ii) age-specific fertility rate assumptions. \ref{sec.a1} reports annual admissions, mean LOS per site, and projected births for the Calgary zone from 2022 to 2030.

In Eq. \ref{eq:3}, we set $\eta=1$ and $\psi=1$, to assume that NICU admissions scale proportionally with births. This choice is consistent with the relatively stable admission-to-birth ratios observed in our historical data. Setting $\psi=1$ treats births as the primary demand driver and excludes additional structural drift. However, alternative $\psi$ values could represent persistent shifts in admission practices or referral patterns that are not explained by birth counts alone.

Table~\ref{tab:projection2023} reports projected bed requirements from the retrospective validation exercise. We summarize projections using medians and IQR, as well as means and standard deviations. The $B_{\text{average}}$ values are fixed, as they depend on deterministic yearly admissions $\hat{A}_y$ and mean LOS over $Y_h^\nu$. Other strategies vary due to resampled within-year arrival and LOS patterns. We compare these projections with retrospective capacity estimates in Table~\ref{tab:bed_estimates_vs_actual}, which were computed using the full historical time series. .

Across sites, $B_{\text{average}}$ remains close to historical values. For Sites 1, 3, 4, and 5, projected medians under $B_{0.05}$ and $B_{0.01}$ align with historical estimates and show narrow IQRs, indicating stability in recent arrival and LOS patterns. For Site 2, projected risk-constrained capacities are slightly lower, as estimates in Table~\ref{tab:bed_estimates_vs_actual} were driven by joint peaks in admissions and LOS. However, the projection in the retrospective validation exercise anchors admissions to recent conditions and resamples within-year patterns, which reduces the influence of extreme historical periods. Overall, the results show that the approach preserves the scale and cross-site ordering of historical bed requirements, while yielding estimates that reflect more recent structural conditions and realistic variability through resampling.

\begin{table}[!ht]
\caption{Projected bed requirements (2022-2023), summarized across $R=300$ runs}
\footnotesize
\centering
\bgroup
\def\arraystretch{0.5}
\begin{tabular}{p{0.4cm}p{0.9cm}p{2.2cm}p{1.6cm}p{2.2cm}p{1.6cm}}
\toprule
Site  & $B_{\text{average}}$
& \multicolumn{2}{c}{$B_{0.05}$} & \multicolumn{2}{c}{$B_{0.01}$} \\
\addlinespace[2pt]
\cmidrule{3-4} \cmidrule{5-6} 
\addlinespace[2pt]
&  & median [IQR] & mean (SD) & median [IQR] & mean (SD) \\
\midrule
1  & 19 & 26 (22, 27) & 25 (3) & 30 (26, 31) & 29 (4) \\
2  & 68 & 74 (71, 77) & 74 (4) & 81 (78, 85) & 81 (5) \\
3  & 14 & 18 (18, 19) & 19 (1) & 22 (22, 23) & 22 (1) \\
4  & 17 & 23 (21, 23) & 22 (1) & 27 (25, 28) & 26 (2) \\
5  & 10 & 15 (14, 16) & 15 (1) & 18 (17, 19) & 18 (1) \\
\bottomrule
\end{tabular}
\egroup
\label{tab:projection2023}
\end{table}

Table~\ref{tab:projection2030} reports projected beds required in 2030 under each strategy. The results show that $B_{0.05}$ and $B_{0.01}$ consistently exceed $B_{\text{average}}$, reflecting additional beds needed to keep overflow probability low. For example, at Site 1 the median $B_{0.05}$ is 26 beds versus 21 under $B_{\text{average}}$, and the median $B_{0.01}$ increases to 31. Across sites, the median increase relative to $B_{\text{average}}$ ranges from 4–6 beds for $B_{0.05}$ and 8–13 beds for $B_{0.01}$. This illustrates how progressively tighter risk thresholds translate into higher capacity targets.

The dispersion across scenarios provides further insight into projection robustness and it varies by site. Larger sites such as Site~2 exhibit wider interquartile ranges (up to 7 beds) and standard deviations (up to 5 beds), whereas smaller sites such as Site~5 show tighter interquartile ranges (up to 2 beds) and standard deviations (up to 1 bed). Within each site, variability generally increases from $B_{0.05}$ to $B_{0.01}$, which reflects the greater sensitivity of conservative planning rules to daily demand fluctuations.

\begin{table}[!ht]
\caption{Projected bed requirements (2024-2030), summarized across $R=300$ runs}
\footnotesize
\centering
\bgroup
\def\arraystretch{0.5}
\begin{tabular}{p{0.4cm}p{0.9cm}p{2.2cm}p{1.6cm}p{2.2cm}p{1.6cm}}
\toprule
Site  & $B_{\text{average}}$
& \multicolumn{2}{c}{$B_{0.05}$} & \multicolumn{2}{c}{$B_{0.01}$}\\
\addlinespace[2pt]
\cmidrule{3-4} \cmidrule{5-6}
\addlinespace[2pt]
&  & median [IQR] & mean (SD) & median [IQR] & mean (SD) \\
\midrule
1  & 21 & 26 (25, 29) & 27 (4) & 31 (29, 34) & 32 (4) \\
2  & 72 & 79 (75, 83) & 79 (5) & 87 (82, 92) & 87 (5) \\
3  & 16 & 21 (20, 22) & 21 (1) & 25 (24, 26) & 26 (2) \\
4  & 18 & 24 (23, 25) & 24 (2) & 28 (26, 30) & 28 (2) \\
5  & 12 & 17 (17, 19) & 18 (1) & 21 (20, 23) & 21 (2) \\
\bottomrule
\end{tabular}
\egroup
\label{tab:projection2030}
\end{table}

\subsection{Implications for Capacity Planning}

In what follows, we summarize the main findings in an application-oriented manner and discuss their relevance for long-term capacity planning under non-stationary demand.

\begin{myenum}

\item \textit{Average-based utilization rules or fixed utilization targets may underestimate overflow risk.}

Planning based on average occupancy or fixed utilization targets, including the 85\% benchmark, ignores temporal variability in admissions and LOS. Because both are time-varying, rules relying on long-run averages may underestimate required beds. Furthermore, sustaining utilization above a fixed threshold such as 85\% in a non-stationary environment increases the risk of overflow, delayed admissions, or service strain. Variation across sites further suggests that capacity strategies should be tailored to local demand characteristics rather than applied uniformly across institutions.

\item \textit{Resiliency-based thresholds provide a risk-explicit framework for capacity determination.}

Overflow-constrained strategies set bed levels by imposing an explicit bound on the probability that demand exceeds capacity. This shifts planning from utilization heuristics to a probabilistic performance criterion that quantifies congestion risk. By expressing decisions through exceedance probabilities, alternative thresholds can be evaluated based on their resiliency implications. Such an approach provides a quantitative basis for selecting bed levels consistent with institutional risk tolerance and observed variability in admissions and LOS, rather than relying only on average occupancy targets.

\item \textit{The efficiency–resiliency trade-off is inherent in non-stationary ICU systems.}

Because admissions and LOS vary over time, increasing bed capacity reduces overload risk but lowers average utilization, whereas maintaining high utilization increases congestion exposure. This trade-off reflects variability in \(\lambda_t\) and \(\mu_t\) and persists across sites and projections. It can become more pronounced in non-stationary systems, where both admissions and LOS vary over time, making it more difficult to simultaneously maintain high utilization and protection against overflow. More strict thresholds such as \(B_{0.01}\) offer stronger protection against exceedance but further reduce utilization, while \(B_{0.05}\) may offer a balance between routine performance and surge resilience. The appropriate threshold depends on institutional risk tolerance and local demand variability. Expanding permanent bed capacity to address surges reduces average utilization and can leave resources underused during routine periods, whereas tuning capacity primarily for efficiency increases exposure to prolonged over-occupancy. One potential response is to supplement fixed infrastructure with flexible components such as temporary surge beds, cross-trained staff, or shared arrangements with other units. While challenging in neonatal settings, such mechanisms may mitigate the structural efficiency–resiliency trade-off observed across sites.

\item \textit{Redistribution may mask structural imbalances in demand across sites.}

In a multi-site NICU network, inter-site transfers can ease local congestion but may conceal persistent differences in underlying demand. When high-demand sites are routinely relieved through transfers, historical bed allocations may appear balanced despite unbalanced intrinsic demand. By removing redistribution in planning, we estimate the capacity each site would need to serve its own demand. In our setting, we observe that one site would require substantially higher capacity than its current allocation suggests, which indicates regional imbalance. If adjusting permanent capacity at high-demand sites is constrained, continued reliance on transfers becomes a structural mechanism for redistributing demand, which can be incorporated explicitly into regional capacity planning rather than treated solely as an operational response.

\item \textit{LOS dispersion influences risk-sensitive capacity estimates.}

Required bed levels under resiliency-based planning depend on both the mean and dispersion of the LOS distribution. With mean LOS fixed, changes in variance alter the degree of temporal overlap among patients and therefore affect peak occupancy formation. Greater dispersion spreads discharges over time and moderates peaks, whereas lower variance concentrates departures and increases short-term congestion risk.

\item \textit{Capacity projections should account for multiple sources of uncertainty.}

Projected bed requirements are affected both by parameter uncertainty in estimated admission rates and LOS, as well as structural uncertainty in demand fluctuations beyond historical patterns. Demographic shifts, episodic surges, or clusters of prolonged stays may produce realizations that differ from the past patterns. Interpreting projections as ranges rather than point estimates, regularly updating inputs, and evaluating alternative fluctuation scenarios can provide a more stable basis for long-term planning under evolving demand conditions.

\end{myenum}
\section{Discussion} \label{Discussion}

This study develops a data-driven framework to estimate time-varying ICU bed occupancy under non-stationary demand using an $M_t/G_t/\infty$ structure informed by empirical arrival and LOS processes. Rather than relying on steady-state or time-aggregated averages that ignore temporal heterogeneity, the framework models temporal variability in admissions and service durations and translates occupancy trajectories into overflow-constrained capacity levels.

The main methodological contribution is integrating STL decomposition, parametric LOS modeling, and occupancy estimation within a unified, tractable pipeline. The framework supports scenario-based evaluation of resiliency thresholds by translating projected occupancy into exceedance probabilities. A key strength of the $M_t/G_t/\infty$ structure is its ability to assess forward-looking scenarios by adjusting time-varying admissions or LOS distributions to evaluate changes in practice, policy, or demand. It also includes a births-driven projection component linking demographic forecasts with site-level admission patterns, enabling planning that reflects expected population growth rather than relying only on historical data.

Our results show clear differences between utilization-based and overflow-constrained strategies in non-stationary settings. Capacity based solely on average occupancy may underestimate exceedance risk by ignoring temporal variation. In contrast, probabilistic thresholds determine bed levels using explicit resiliency targets. We also find that variability in arrivals and LOS, including dispersion effects, influences the capacity required to meet these targets.

Several limitations should be acknowledged. First, the model operates at daily resolution and does not capture intra-day fluctuations in admissions or discharges. Second, STL smoothing may lag abrupt demand shifts. Third, the LOS component assumes fixed shape parameters, which may not fully capture evolving clinical dynamics. Furthermore, in our setting, overflow probabilities rely on a Poisson approximation, which may understate tail risk under correlated arrivals. 

A further limitation is potential error in estimating the LOS distribution, especially when bias leads to systematic over- or underestimation and affects planning resiliency. Rare extreme cases (e.g., year-long stays) are also difficult to estimate due to sparse tail data, limiting accurate estimation of extreme occupancy levels. This warrants caution and may justify conservative adjustments for tail uncertainty. Stratifying patients into more homogeneous subgroups (e.g., by diagnosis or LOS profile) could improve fit, but small subgroup sizes—particularly among long-stay patients—can make parameter estimation and scenario analysis unstable and introduce additional uncertainty. Finally, our framework does not model staffing constraints, routing, or endogenous discharge under congestion. Mortality is embedded within LOS rather than treated as a separate exit process, and thus distinguishing discharge and death may improve interpretability in settings where they follow distinct dynamics.

Several extensions could strengthen our framework. Integration with an interactive decision-support platform would enable real-time scenario evaluation. Short-term machine learning forecasts of time-varying arrival rates and LOS moments could be combined with the queueing structure to create hybrid predictive-planning pipelines. In addition, online updating mechanisms may further improve responsiveness to changing demand conditions.

Future research may incorporate explicit modeling of transfers and routing to examine coordinated regional capacity design. Although our empirical analysis uses neonatal data, the modeling structure is not population-specific and can be applied to other ICU settings with appropriate characterization of arrivals and LOS dynamics. In environments where congestion induces state dependence in LOS, the service component would need adjustment, while the occupancy estimation formulation remains applicable. The framework can also extend to other hospital units (e.g., surgical recovery) and geographically dispersed systems, potentially supporting coordinated multi-region planning. Incorporating covariates such as weather, staffing, or policy changes may further enhance predictive performance and adaptive planning, which we leave for future work.

\section*{Ethics Approval}

Our study has received ethical approval from the University of Calgary’s Conjoint Health Research Ethics Board (CHREB) under ethics ID REB24-0800. All data are collected in accordance with Alberta Health Services (AHS) data governance protocols, with direct identifiers removed prior to researcher access. Only de-identified records of NICU admissions, discharges, and intra-hospital transfers are used for analysis. Data confidentiality and privacy are maintained in compliance with AHS and institutional guidelines.

\section*{Acknowledgments}

We sincerely thank Sharon Zhang and Bing Li at Alberta Health Services for their support in coordinating and providing access to the data, which significantly contributed to the development of this study. 

\section*{Funding}

This work was supported by the Natural Sciences and Engineering Research Council of Canada (NSERC) Discovery Grant Program. We also acknowledge funding support from the University of Calgary Transdisciplinary Scholarship Connector Grant, with Dr. Na Li as Principal Investigator.

\clearpage
\appendix

% \appendix
\section{LOS Tail Comparisons}\label{sec.a2}

\begin{figure}[!ht]
\centering

\begin{subfigure}[t]{0.48\textwidth}
\centering
\includegraphics[width=\textwidth]{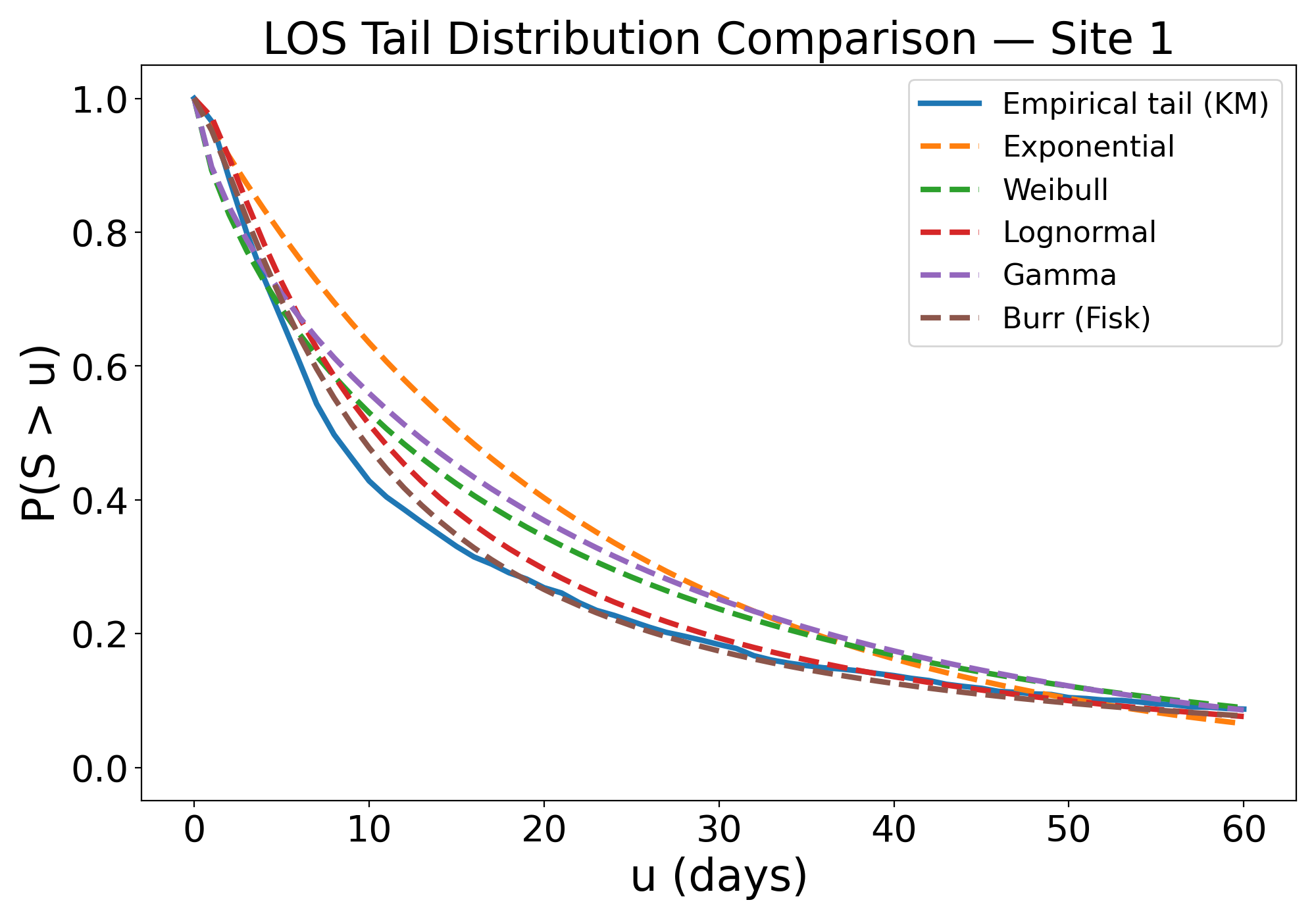}
\caption{Site 1}
\end{subfigure}\hspace{0.02\textwidth}%
\begin{subfigure}[t]{0.48\textwidth}
\centering
\includegraphics[width=\textwidth]{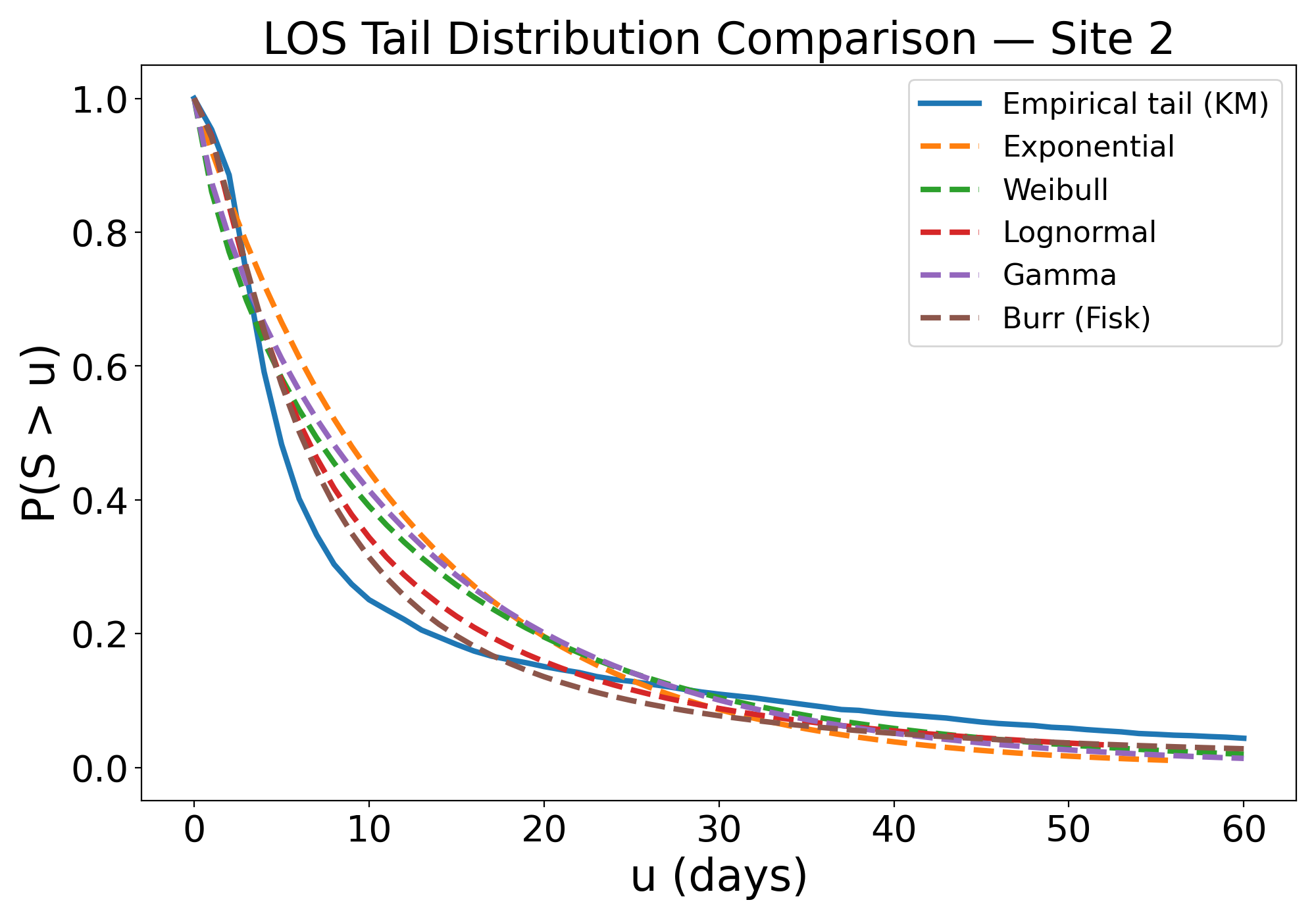}
\caption{Site 2}
\end{subfigure}

\begin{subfigure}[t]{0.48\textwidth}
\centering
\includegraphics[width=\textwidth]{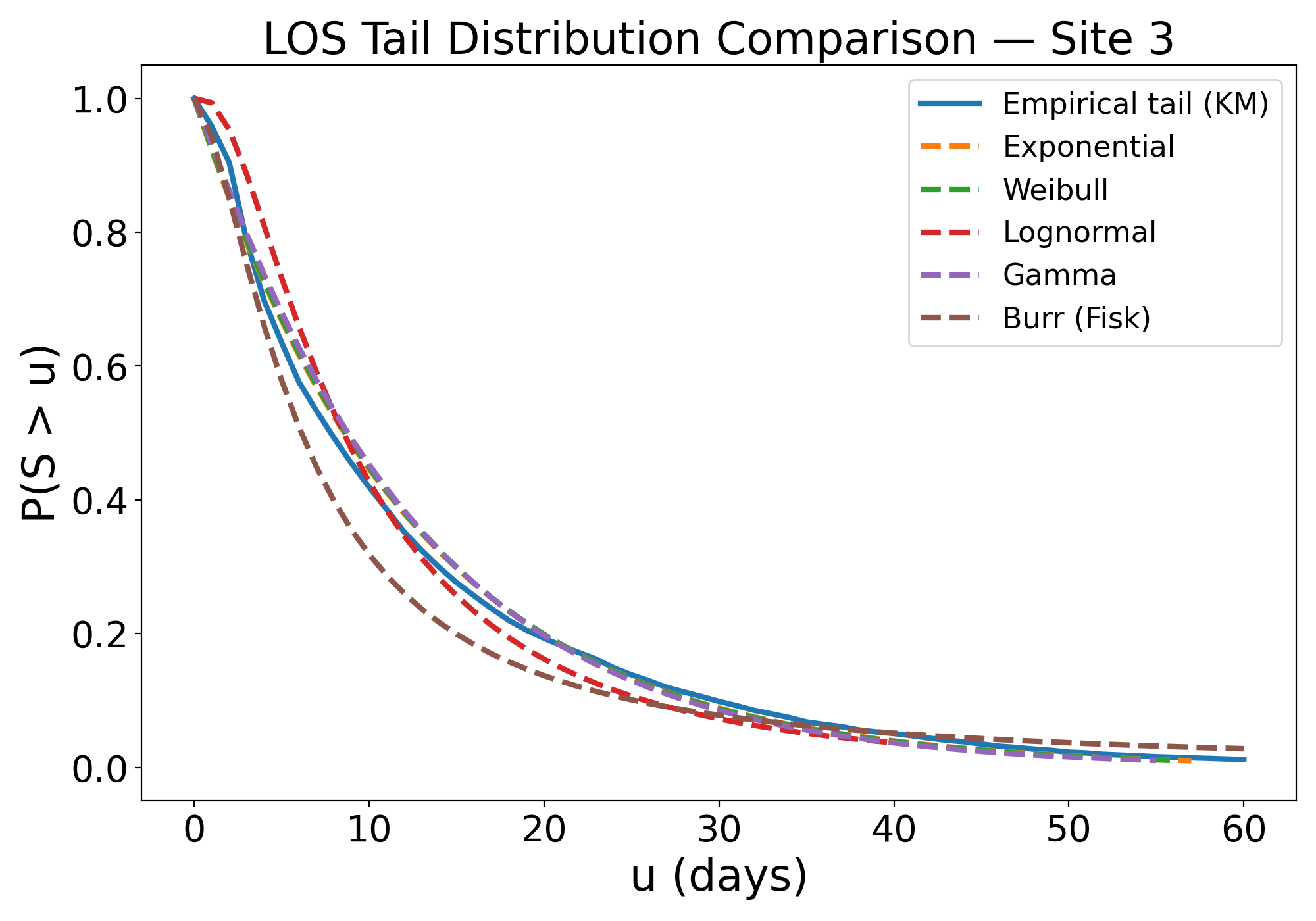}
\caption{Site 3}
\end{subfigure}\hspace{0.02\textwidth}%
\begin{subfigure}[t]{0.48\textwidth}
\centering
\includegraphics[width=\textwidth]{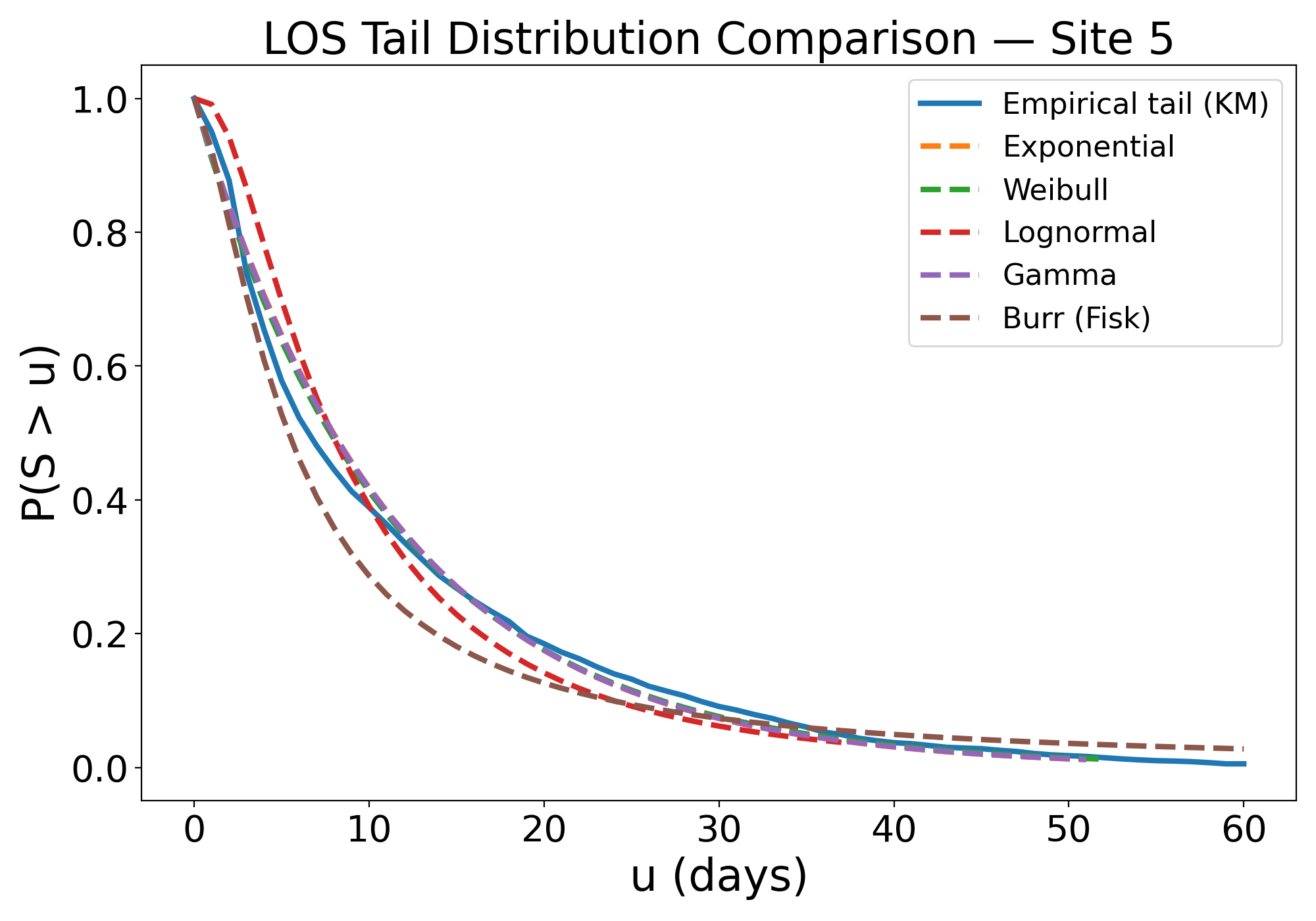}
\caption{Site 5}
\end{subfigure}

\caption{Comparison of empirical LOS tail probabilities $\mathbb{P}(S > u)$ with candidate parametric distributions at the remaining sites}
\label{fig:tail_comparisons2}
\end{figure}

\section{Annual Admissions and Mean LOS (Historical) and Projected Births}\label{sec.a1}

Table \ref{tab:admissions_los} reports annual total admissions and mean LOS for each site during the historical period from 2016 to 2023. Admissions are consistently highest at Site 2, which records nearly 1,000 admissions each year, while Site 1 shows the lowest volumes. Mean LOS varies more widely across sites, with Site 1 displaying the longest stays on average and Sites 2 and 5 the shortest. Year-to-year fluctuations are evident, but no strong upward or downward trend is observed in admissions, whereas LOS values remain relatively stable within each site. The projected births for the Calgary zone, provided by the Alberta Interactive Health Data repository \citep{albertadata} increase steadily from 18,741 in 2022 to 21,049 in 2030, with annual projections of 19,067 (2023), 19,337 (2024), 19,569 (2025), 19,822 (2026), 20,087 (2027), 20,383 (2028), and 20,703 (2029).

\begin{table}[!ht]
\centering
\caption{Historical admissions and mean LOS per site}
\label{tab:admissions_los}
\begin{tabular}{lccccc}
\toprule
Year & Site 1 & Site 2 & Site 3 & Site 4 & Site 5  \\
\midrule
2016 & 218 / 21.20 & 823 / 11.58 & 585 / 11.03 & 463 / 13.89 & 343 / 10.29 \\
2017 & 299 / 20.08 & 1094 / 20.70 & 527 / 8.64 & 449 / 11.05 & 334 / 8.60 \\
2018 & 330 / 18.24 & 1158 / 21.92 & 492 / 8.39 & 452 / 9.96 & 344 / 8.76 \\
2019 & 265 / 25.82 & 1039 / 21.81 & 466 / 9.99 & 514 / 10.98 & 356 / 9.27 \\
2020 & 250 / 23.57 & 989 / 20.92 & 397 / 9.17 & 491 / 9.68 & 305 / 8.47 \\
2021 & 229 / 26.79 & 996 / 22.70 & 438 / 8.95 & 481 / 9.53 & 332 / 8.86 \\
2022 & 250 / 21.87 & 954 / 21.97 & 454 / 8.60 & 504 / 8.99 & 421 / 8.17 \\
2023 & 314 / 20.50 & 990 / 23.56 & 532 / 9.40 & 539 / 9.07 & 393 / 8.46 \\
\bottomrule
\multicolumn{6}{l}{\footnotesize Values shown as ``yearly admissions / mean LOS''} \\
\end{tabular}
\end{table}

\vspace{-0.95cm}
\bibliographystyle{elsarticle-num} 
\bibliography{references}

\end{document}